\documentclass[12pt,a4paper]{article}
\usepackage{cite}
\usepackage{amsmath,amssymb,graphicx}
\usepackage{float}

\numberwithin{equation}{section}
\allowdisplaybreaks[3]

\newcommand{\al}{\alpha}
\newcommand{\ep}{\epsilon}
\newcommand{\vep}{\varepsilon}

\newcommand{\la}{\lambda}
\newcommand{\La}{\Lambda}

\newcommand{\msc}[1]{\mbox{\scriptsize #1}}
\newcommand{\dsp}{\displaystyle}
\newcommand{\scs}[1]{{\scriptstyle #1}}

\newcommand{\bm}[1]{\mbox{\boldmath ${#1}$}}

\newcommand{\bc}{\Bbb C}
\newcommand{\br}{\Bbb R}
\newcommand{\bz}{\Bbb Z}

\newcommand{\bh}{\Bbb H}

\renewcommand{\-}{{\bf -1}}

\newcommand{\cT}{{\cal T}}

\newcommand{\cN}{{\cal N}}
\newcommand{\cM}{{\cal M}}
\newcommand{\cF}{{\cal F}}

\newcommand{\cS}{{\cal S}}

\newcommand{\cC}{{\cal C}}

\newcommand{\cH}{{\cal H}}

\newcommand{\cK}{{\cal K}}

\newcommand{\tL}{\tilde{L}}

\newcommand{\ti}{\tilde{i}}

\newcommand{\tchi}{\tilde{\chi}}

\newcommand{\hZ}{\widehat{Z}}

\newcommand{\ket}[1]{{\left|#1\right\rangle}}

\newcommand{\Th}[2]{\Theta_{#1,#2}}
\renewcommand{\th}{{\theta}}
\newcommand{\tTh}[2]{\widetilde{\Theta}_{#1,#2}}

\newcommand{\tr}{\mbox{Tr}}
\renewcommand{\mod}{\mbox{mod}}

\newcommand{\nn}{\nonumber\\}

\newcommand{\del}{\partial}

\newcommand{\any}{{}^{\forall}}

\newcommand{\bigbox}[2]{{\scs{#1}} \hspace{0.3mm}  \raise-1mm\hbox{$ \underset{#2} {\scalebox{1.9}{\mbox{$\Box$}}} $}}

\newcommand{\eqn}[1]{(\ref{#1})}

\begin{document}

\begin{titlepage}
 \
 \renewcommand{\thefootnote}{\fnsymbol{footnote}}
 \font\csc=cmcsc10 scaled\magstep1
 {\baselineskip=16pt
  \hfill
}

 \baselineskip=20pt
\vskip 1cm
 
\begin{center}

{\bf \Large 


Notes on Vanishing Cosmological Constant

without Bose-Fermi Cancellation

} 

 \vskip 1.2cm

\noindent{ \large Yuji Satoh}\footnote{\sf ysatoh@u-fukui.ac.jp},

\medskip

{\it Department of Applied Physics, University of Fukui\\
Bunkyo 3-9-1, Fukui 910-8507,  Japan}

\vskip 8mm


\noindent{ \large Yuji Sugawara}\footnote{\sf ysugawa@se.ritsumei.ac.jp},

\medskip

 {\it Department of Physical Sciences, 
 College of Science and Engineering, \\ 
Ritsumeikan University,  
Shiga 525-8577, Japan}

\end{center}

\bigskip

\begin{abstract}

In this article we discuss how one can systematically construct the {\em point particle theories}
that realize the vanishing one-loop cosmological constant without the bose-fermi cancellation. 
Our construction is based on the asymmetric (or non-geometric) orbifolds of supersymmetric string vacua.
Using the building blocks of their partition functions and their modular properties, we construct the theories which would be naturally identified 
with certain point particle theories including infinite mass spectra, 
but {\em not\/} with string vacua. 
They are obviously non-supersymmetric due to the mismatch of
the  bosonic and fermionic degrees of freedom at each mass level. 
Nevertheless, it is found that the one-loop cosmological constant vanishes, 
after removing the parameter effectively playing the role of  the UV cut-off.
As concrete examples we demonstrate the constructions of the models based on the toroidal asymmetric orbifolds 
with the Lie algebra lattices (Englert-Neveu lattices)
by making use of the analysis given in \cite{SSLie}.

\end{abstract}

\setcounter{footnote}{0}
\renewcommand{\thefootnote}{\arabic{footnote}}

\end{titlepage}

\baselineskip 18pt

\vskip2cm 
\newpage


\section{Introduction}

String theories on the asymmetric orbifolds, which should correspond to some non-geometric backgrounds,  
have  interesting aspects.  
Among others
it is  remarkable that the string vacua with  the vanishing cosmological constant 
can be realized without the help of
unbroken SUSY.  
Such studies 
have been initiated 
in type II string theories
by \cite{Kachru1,Kachru2,Kachru3} 
based on some non-abelian orbifolds, 
followed by studies {\em e.g.} in 
\cite{Harvey,Shiu-Tye,Blumenhagen:1998uf,Angelantonj:1999gm,Antoniadis,Aoki:2003sy}. 
More recently, several non-SUSY vacua with this property have been constructed  
as asymmetric orbifolds \cite{Narain:1986qm} by simpler cyclic groups in \cite{SSW,SWada}.
In heterotic string theories, on the other hand, there have been many studies on the string vacua with 
the cosmological constant exponentially suppressed with respect to some moduli 
(for instance, the radii of tori of compactifications as given in \cite{SS}) 
in  \cite{IT,Harvey}, and more recently, {\em e.g.} in \cite{Blaszczyk:2014qoa,Angelantonj:2014dia,Faraggi:2014eoa,Abel:2015oxa,
Kounnas:2015yrc,Abel:2017rch,ItoyamaN} closely related with the model buildings in string phenomenology. 


However, in all of these models, the vanishing (or exponentially suppressed) cosmological constant 
is eventually achieved by the bose-fermi cancellation, 
even though the space-time SUSY is broken. 
Toward realistic particle content,
 it would be more preferable to realize the vanishing (or very small) cosmological constant without the bose-fermi cancellation. 
In string theory this means that 
\begin{equation}
Z(\tau) \not\equiv 0, ~~ \mbox{but,} ~~  \La \equiv \int_{\cF} \frac{d^2 \tau}{\tau_2^2}\, Z(\tau) =0 
\hspace{1cm} \left(\tau \equiv \tau_1 + i\tau_2 \in \bh^+ \right) ,
\label{vanising cc string}
\end{equation}
at the one-loop level (at least).
Here, 
$Z(\tau)$ is the torus partition function, and 
$\cF$ denotes the familiar fundamental region of 
the full modular group $\Gamma(1) \equiv SL(2, \bz)$, in other words, the moduli space of the world-sheet torus,
\begin{equation}
\cF:= \left\{ \tau \in \bh^+ ~:~ -\frac{1}{2} \leq \tau_1 < \frac{1}{2} , ~ \left|\tau \right| \geq 1 \right\}.
\label{cF}
\end{equation}
Unfortunately it has been known to be very difficult to 
find the string vacua with this property. 
An elaborated mechanism has been proposed 
in \cite{Moore} 
 to achieve them based on some modular symmetry argument 
by the Atkin-Lehner involution. 
However, concrete constructions of such string vacua that are physically consistent 
along this line are still very difficult 
\cite{Taylor,Balog:1988dt,Dienes}. 


In this paper, instead of searching the string vacua, we discuss how one can systematically construct 
the {\em particle theories} 
with the preferable property mentioned above, that is, the vanishing one-loop cosmological constant without the bose-fermi cancellation, 
\begin{equation}
Z_{\msc{particle}}(\ell) \not\equiv 0, 
~~ \mbox{but,} ~~  \La \equiv \lim_{\vep \, \rightarrow\, +0} \int_{\vep}^{\infty} \frac{d \ell}{\ell}\, Z_{\msc{particle}}(\ell) =0,
\label{vanishing cc particle}
\end{equation}
where $\ell $ denotes the Schwinger parameter (modulus of the world-line circle) 
and $\vep$ is 
the UV cut-off, 
based on some non-geometric orbifolds of superstring vacua. 
Indeed,
as will be demonstrated,
 $Z_{\msc{particle}}(\ell) $ given here originates from
the partition sum only of the {\em untwisted sector} of 
the relevant orbifold models
rather than the total one,
which is 
only invariant under the modular T-transformations $\tau \, \rightarrow \, \tau + n$ ($\any n \in \bz$). 
Consequently, the natural integration region of modulus $\tau$ should be 
the `strip region' 
\begin{equation}
\cS:= \left\{ \tau \in \bh^+ ~:~ -\frac{1}{2} \leq \tau_1 < \frac{1}{2} \right\},
\label{cS}
\end{equation}
in place of $\cF$. We then naturally obtain $Z_{\msc{particle}}(\ell)$ by identifying $\tau_2 $ with $ \ell $ and integrating $\tau_1$ out, which just amounts to 
imposing the level-matching condition.
In the main part of this article,
we will clarify how to construct the appropriate non-geometric orbifold models 
whose untwisted sector 
yields the particle theories with the desired property 
\eqn{vanishing cc particle}.
Of course, as working with particle theories (the integration region should be $\cS$ rather than $\cF$), 
one would have to introduce the UV cut-off as in \eqn{vanishing cc particle}.
We shall actually adopt a generalization of the Scherk-Schwarz type spontaneous SUSY breaking \cite{SS}, 
in which the compactification radius  
turns out to effectively play the role of the  UV cut-off.

~


This paper is organized as follows:
In section 2, we discuss how one can obtain
the particle theories with the property \eqn{vanishing cc particle} from rather general setups of orbifolds. 
We propose the conditions that should be satisfied by the wanted models,
and prove that  they are indeed  sufficient to  realize \eqn{vanishing cc particle}. 
%
In section 3, we demonstrate  a systematic construction of the concrete models satisfying these conditions based on the type II and heterotic 
string vacua compactified on toroidal asymmetric orbifolds.
Especially, the models we focus on are the ones  
given in \cite{SSLie}, with some extensions and refinements included. 
The main part of section 3 would look a little  technical, but, 
we also present simple examples in order to clarify the general features of our models. 
%
In section 4, 
we conclude with several comments and discussion. 


~


\section{How to Achieve Vanishing Cosmological Constant}
\label{sec 2}

In this section we discuss how to obtain the particle theories from general setups of orbifolds which possess the property \eqn{vanishing cc particle}.
Among others, we clarify the conditions that should be satisfied by the relevant orbifold models.


We start with a generic  superstring vacuum with unbroken space-time SUSY, which we
tentatively denote as `$\cM_0$'.
More precisely, we assume that $\cM_0$ is defined as the following background,
$$
\cM_0 \cong \br^{D-1,1} \times \cK \times S^1_R,
$$
where  
$S^1_R$ denotes the circle with radius $R \sqrt{\al'}$. 
The `internal sector', which we also denote by $\cK$, 
only has a discrete spectrum.

Let $g$ be an order $N_0$ automorphism that acts on  
$\cK$ as well as the world-sheet fermions  
and does not commute 
with any space-time supercharges. 
We further assume the existence of the modular invariant partition function for
the $\bz_N$-orbifold of $\br^{D-1,1} \times \cK \left( \otimes [\mbox{fermions}]\right) $ defined by the $g$-action, 
where $N$ is a certain multiple of $N_0$ \footnote{We allow the cases of $N \neq N_0$, which typically happens for the {\em asymmetric\/} orbifolds due to 
the existence of non-trivial phase factors in the twisted sectors
 (see e.g.\cite{Aoki:2004sm,SSW,Harvey:2017rko}).
We will actually work with the orbifolds of this type in section \ref{sec:simple models}.
}. 
This is written in the form, 
\begin{equation}
Z_{\msc{orb}}(\tau) \equiv \frac{1}{N} \sum_{a, b \in \bz_N}\, Z_{(a,b)}(\tau),
\label{ZN-orb}
\end{equation}
where
$a,b \in \bz_N$ label the twistings 
along the spatial and the temporal direction respectively 
with respect to the orbifold action $g$ in the world-sheet torus.
In other words, one may write
\begin{equation}
Z_{(a,b)}(\tau) \equiv \tr_{\cH_{a}
} \left[g^{b}\, 
q^{L_0-\frac{c_L}{24}} \overline{q^{\tL_0-\frac{c_R}{24}}}  \right],
\label{Z al beta def}
\end{equation}
where $\cH_{a}$ denotes the Hilbert space of the  twisted sector 
associated to the $g^{a}$ action. $Z_{(a,b)}(\tau)$ should possess the periodicity
\begin{equation}
Z_{(a+N,b)}(\tau) = Z_{(a,b+N)}(\tau) = Z_{(a,b)}(\tau),
\end{equation}
for consistency. 
Moreover, one should require 
the `modular covariance';
\begin{align}
& 
\left. Z_{(a,b)}(\tau) \right|_S \left( \equiv Z_{(a,b)} \left(-\frac{1}{\tau} \right) \right)
= Z_{(b,-a)}(\tau),
\label{mod cov S}
\\
& \left. Z_{(a,b)}(\tau)\right|_T \left(\equiv Z_{(a,b)}\left(\tau+1 \right) \right)
= Z_{(a, a+b)}(\tau),
\label{mod cov T}
\end{align}
in the standard fashion.


Now,
setting $R = N \ep $ with a small positive number $\ep$, let us consider the orbifold of $\cM_0\equiv \br^{D-1,1} \times \cK \times S^1_{N\ep}$ defined by the operator
\begin{equation}
\bm{g} := g \otimes \cT_{2\pi \ep}, 
\label{bm g}
\end{equation}
where $\cT_{2\pi \ep}$ denotes the translation 
$
X \, \rightarrow \, X + 2\pi \ep \sqrt{\al'} \, 
$
along $S^1_{N\ep}$.
One may regard this as a generalization of the Scherk-Schwarz type compactification \cite{SS}. 
Thus, the models we will construct are actually those with a  spontaneously broken SUSY,  in which 
the relevant SUSY is recovered when taking the `Scherk-Schwarz radius' $\ep$ to be infinity. 
We note that the vacuum energy is not lifted as $\ep$ is varied, since 
it parameterizes a flat direction.
We shall later take the $\ep \, \rightarrow \, +0$ limit. 
The torus partition function of the $\bm{g}$-orbifold is written as   
\begin{align}
\bm{Z}_{\msc{orb}}(\tau) 
& = 
\frac{1}{N} \sum_{a, b \in \bz_N}\,
Z_{(a,b)}(\tau) \, \frac{N \ep}{\sqrt{\tau_2} \left|\eta(\tau)\right|^2}
\sum_{w,m\in \bz}\, e^{- \frac{\pi \left(N \ep\right)^2}{\tau_2}\left| \left(w + \frac{a}{N}\right) \tau+ \left(m + \frac{b}{N}\right)\right|^2}
\nn
& \equiv  \frac{ \ep}{\sqrt{\tau_2} \left|\eta(\tau)\right|^2} \sum_{w,m \in \bz}\, Z_{(w,m)}(\tau) \, e^{- \frac{\pi \ep^2}{\tau_2} \left|w \tau + m \right|^2} ,
\label{total ZN-orb}
\end{align}
which is manifestly modular invariant.


Then, 
we define the `theory $\cM [\ep]$' as that contains 
the mass spectrum
read off from the level-matched
sector of the following  partition function 
\begin{equation}
\bm{Z}_{\cM[\ep]} (\tau) := \frac{ \ep}{\sqrt{\tau_2} \left|\eta(\tau)\right|^2}  \sum_{m\in \bz}\, Z_{(0,m)}(\tau) \, e^{-\frac{\pi}{\tau_2} \ep^2 m^2}.
\label{ZcM}
\end{equation}
This represents  the contributions to $ \bm{Z}_{\msc{orb}}(\tau) $
for the $\bm{g}$-orbifold
from the untwisted sector
with  {\em no spatial winding number\/} along $S^1_{N\ep}$.
The exponential factor $e^{-\frac{\pi}{\tau_2} \ep^2 m^2}$ 
makes 
the infinite summation 
over $m\in \bz$ converge absolutely as long as $\ep>0$. 
One can  read off the mass spectrum from the partition function \eqn{ZcM} by making the Poisson resummation for the temporal winding $m \in \bz$. 
The overall factor $\ep$ is absorbed after this resummation.
Since we have 
$
\cT_{2\pi \ep} \equiv \exp \left[- i 2\pi \ep \sqrt{\al'} P \right],
$
where $P$ denotes the KK momentum operator along $S^1_{N\ep}$, 
the states with the $g$-eigenvalue $e^{2\pi i \frac{r}{N}}$ have to possess the KK momenta $n \equiv r ~ (\mod\, N)$, implying that these excitations 
at least have masses of order 
$\dsp \sim  \frac{1}{\ep} M_s \left(\equiv \frac{1}{\ep \sqrt{\al'}} \right)$
except for the $r=0$ case. 
In particular all the massless states in $\cM[\ep]$ are lying in $\dsp Z_0(\tau)\equiv \frac{1}{N} \sum_{b \in \bz_N}\, Z_{(0,b)}(\tau)$.


We emphasize that $\bm{Z}_{\cM[\ep]}(\tau)$  
is {\em not\/}  invariant under the full modular group.
This is invariant only under the modular T-transformations. 
Thus, we cannot adopt the usual fundamental region $\cF$
of $\bh^+/\Gamma(1)$ given in \eqn{cF}
for the integration of the modulus $\tau$. 
In other words, the theory $\cM[\ep]$ is {\em not\/} identified with
some string vacuum, although we have an infinite number of mass spectrum.
It is rather natural to take the `strip region' $\cS$ defined in \eqn{cS}
as the appropriate integration region,
which is identified as the fundamental region for  
$\bh^+/\Gamma_0(\infty) \equiv \bh^+/\left\langle T \right\rangle$.
We note that the factor $e^{-\frac{\pi}{\tau_2} \ep^2 m^2}$  for $m\neq 0$ effectively truncates the UV region $\tau_2\, \sim \, +0$,  which makes the moduli 
integral well-defined, even if taking $\cS$ instead of $\cF$. In other words the parameter $\ep$ plays the role of  the UV cut-off.

Let us recall the familiar relationship between the one-loop cosmological constants of closed string theory 
and of point particle theory (see {\em e.g.} 
section
7.3 of \cite{Pol book 1}). 
We assume the bosonic (fermionic) mass spectrum $\left\{ m_i \right\}$, $i\in \cH_B$ ($i \in \cH_F$).
The one-loop cosmological constant of the particle theory on the space-time with $D$-dimensional non-compact directions
is schematically expressed in terms of 
the summation of path-integrals over a world-line circle
(with modulus $\ell$); 
\begin{align}
\La_{\msc{particle}} & = \frac{1}{V_{D}}
\left[ \sum_{i\in \cH_{\msc{B}}}\, - \sum_{i\in \cH_{\msc{F}}}\,\right] \, Z_{S^1}(m_i^2)
\nn
& \equiv  \left[ \sum_{i\in \cH_{\msc{B}}}\, - \sum_{i\in \cH_{\msc{F}}}\,\right] 
\, \int_0^{\infty}\frac{d\ell}{\ell}\, \int \frac{d^D p_i}{(2\pi)^D}\, e^{-\ell \frac{\al'}{2}   \left[  p_i^2 + m_i^2 \right]},
\label{La general particle}
\end{align}
where $p_i$ are the zero-mode momenta along $\br^{D-1,1}$ and $V_D$ denotes the volume factor. 
This is naturally compared with that of string theory. 
For instance,  consider the type II string on 
the $\br^{D-1,1} \times 
[\mbox{internal sector}]
$,
where the `internal sector' only includes the discrete spectrum. 
The 1-loop cosmological constant is written schematically as 
\begin{align}
\La_{\msc{string}} & = \frac{1}{V_{D}}  \int_{\cF} \frac{d^2\tau}{\tau^2_2}\, Z(\tau)  ,
\label{La general string}
\end{align}
where 
\begin{equation}
Z(\tau) \equiv \frac{V_D}{\tau_2^{\frac{D-2}{2}}}\,
 \sum_{(i, \ti) \in \cH_{\perp}} \, D(h_i, \tilde{h}_{\ti} ) \, 
q^{h_i-\frac{1}{2}} \overline{q^{\tilde{h}_{\ti}-\frac{1}{2}}} 
\hspace{1cm}\left(q\equiv e^{2\pi i \tau}\right) ,
\end{equation}
denotes the modular invariant partition function on the world-sheet torus\footnote
  {`$(i,\ti) \in \cH_{\perp}$' indicates the transverse degrees of freedom 
in 
$ \br^{D-2} \times [\mbox{internal sector}]$, 
and the summation is only taken over the discrete 
spectrum of conformal weights.  The Gaussian integral of zero-mode momenta along the transverse direction $\br^{D-2}$
just yields the factor $\tau_2^{- \frac{D-2}{2}}$, while one more factor $\tau_2^{-1}$ due to the  temporal and longitudinal 
zero-mode integrals are
incorporated into the modular invariant measure $d^2\tau/\tau_2^2$.
}.
The coefficients of `degeneracy'   
$D(h_i, \tilde{h}_{\ti})$ are positive (negative) integers for the bosonic (fermionic) states.

The correspondence between \eqn{La general particle} and \eqn{La general string} is  clear, 
if we rewrite 
$\dsp \tau \equiv \frac{\th+ i\ell}{2\pi}$ and identify the imaginary part 
$\ell$ with the circle modulus (Schwinger parameter). 
Indeed, we obtain 
{\em by formally replacing the integration region $\cF$ with $\cS$};
\begin{align}
\hspace{-5mm}
\La_{\msc{string}} \left[\mbox{$\cF$ replaced with $\cS$}\right]
& \equiv  \int_{\cS} \frac{d^2\tau}{\tau^2_2}\, \frac{1}{\tau_2^{\frac{D-2}{2}}}
\, \sum_{i,\ti}
\, D(h_i, \tilde{h}_{\ti} ) \, q^{h_i-\frac{1}{2}} 
\overline{q^{\tilde{h}_{\ti}-\frac{1}{2}}}
\nn
& = \int_0^{\infty}\frac{d\ell}{\ell}\, \int_{-\pi}^{\pi} \frac{d\th}{2\pi} \, \left( 2\pi \al' \ell \right)^{-\frac{D}{2}}
\nn
& \hspace{1cm} \times  \sum_{i, \ti}
\, D(h_i, \tilde{h}_{\ti}) \, 
\exp \left[-\left(h_i+ \tilde{h}_{\ti} -1 \right)\ell + i \theta \left(h_i-\tilde{h}_{\ti} \right)\right],
\label{La string particle}
\end{align}
which easily reduces to  the  expression 
in the form as  \eqn{La general particle}.
In fact, the $\th$-integral just amounts to imposing the level-matching condition 
$h_i= \tilde{h}_{\ti}$, and 
the mass spectrum is identified  as
\begin{equation}
\frac{\al'}{4}m_i^2 = h_i-\frac{1}{2}. 
\end{equation}

Based on these simple arguments, we propose that $\cM[\ep]$ defined above  represents a theory  
with an infinite number of {\em particle spectrum,} 
in which the one-loop cosmological constant is given as 
\begin{align}
\La_{\msc{$\cM[\ep]$}} & \equiv \frac{1}{V_{D}}  \int_{\cS} \frac{d^2\tau}{\tau^2_2}\, \bm{Z}_{\cM[\ep]}(\tau)
\nn
& = \int_0^{\infty} \frac{d\ell}{\ell} \, (2\pi \al' \ell)^{-\frac{D}{2}}\, \sum_{i \in \cH[\cM[\ep]]} \, D(m_i) \, e^{- 2\pi \ell \frac{\al'}{2} m_i^2} ,
\label{La cM}
\end{align}
where the mass spectrum, which is read off from the partition function 
\eqn{ZcM} and by imposing the level-matching condition, is expressed as $\cH[\cM[\ep]]$.  
Again $D(m_i)$ denotes the degeneracy of the mass spectrum including the minus sign for fermions. 
Recall that the  integration region suitable for $\bm{Z}_{\cM[\ep]}(\tau)$ should be $\cS$ rather than $\cF$, as mentioned above.
We can thus naturally identify the partition function $\bm{Z}_{\cM[\ep]}(\tau)$ as describing a spectrum of infinite particles, rather than a string spectrum, 
 without making any formal replacement of 
the integration region of modulus. 

~


Under these preparations  let us exhibit the fundamental requirements to be satisfied by the models we want (some  are already mentioned): 

\begin{description}
\item[(1)] The unorbifolded string vacuum $\cM_0$ has the vanishing partition function, 
\begin{equation}
Z_{\cM_0}(\tau) \equiv \frac{N\ep}{\sqrt{\tau_2} \left|\eta(\tau)\right|^2}\, \sum_{w,m\in \bz}\, e^{- \frac{\pi }{\tau_2} N^2 \ep^2\left|w \tau + m\right|^2} \, Z_{(0,0)}(\tau)
= 0.
\end{equation}
Needless to say, it is enough to start with any supersymmetric vacuum $\cM_0 \equiv \br^{d-1, 1} \times \cK \times S^1_{N\ep}$.


\item[(2)] The partition function $\bm{Z}_{\cM[\ep]}(\tau)$ defined in \eqn{ZcM} does not vanish.
This may imply
\begin{equation}
Z_0(\tau) \equiv \frac{1}{N} \sum_{b\in \bz_N}\, Z_{(0,b)}(\tau) \neq 0. 
\nonumber
\end{equation}
Namely, all the supercharges existing originally in $\cM_0$ are removed by the  $g$-projection.


\item[(3)] The spectrum of the level-matched states appearing in $\bm{Z}_{\cM[\ep]}(\tau)$ is consistent with unitarity. 
Alternatively, we require that  $Z_0(\tau)$ possess  the same property.


\item[(4)] The spectra of the level-matched
states in the sectors $\dsp Z_a(\tau) \equiv \frac{1}{N} \sum_{b\in \bz_N} Z_{(a,b)}(\tau)$ for $\any a \in \bz_N$ do not contain 
tachyons. In other words, we require\footnote
   {The factor $e^{4\pi \tau_2 \frac{1}{24}}$ originates from the factor $\left|\eta\right|^{-2}$ appearing in the $S^1_{N\ep}$-sector. 
   We note that the non-tachyonic behavior for $a \neq 0$ is necessary in order to prove the statement \eqn{La cM 0} (see below (\ref{eval La ep 3})), 
even though only the $a=0$ sector contributes to the spectrum of the theory $\cM[\ep]$.} 
\begin{equation}
\lim_{\tau_2 \rightarrow +\infty} \, e^{4\pi \tau_2 \frac{1}{24}} \left| \int_{-1/2}^{1/2} d\tau_1\, Z_{a}(\tau) \right| 
< + \infty 
\hspace{1cm} (\any a \in \bz_N) .
\label{cond no tachyon}
\end{equation}
(We allow the existence of  level-mismatched tachyons.)

\item[(5)] The orbifold partition function  \eqn{ZN-orb} vanishes.

\end{description}


Now, we  state our main claim: 
under the requirements {\bf (1)} $\sim$ {\bf (5)} given above\footnote
   {In fact, only the requirements {\bf (1)}, {\bf (4)}, {\bf (5)} are sufficient to show the claim \eqn{La cM 0}.}, 
the theory $\cM[\ep]$ ($\ep>0$) 
yields the finite cosmological constant $\La_{\cM[\ep]}$, 
and we further obtain 
\begin{equation}
\lim_{\ep\, \rightarrow \, +0}\,  \La_{\cM[\ep]} =0, ~~~ \mbox{for}~ \any D \geq 1.
\label{La cM 0}
\end{equation}


~

\noindent
{\bf [proof of the claim]}

We first note that $\La_{\cM[\ep]}$ defined in \eqn{La cM} is explicitly written as 
\begin{align}
\La_{\msc{$\cM[\ep]$}} 
= \frac{1}{V_D} \int_{\cS} \frac{d^2\tau}{\tau^2_2}\, \frac{\ep}{\sqrt{\tau_2} \left|\eta(\tau) \right|^2}\, 
\sum_{m\in \bz}\, Z_{(0,m)}(\tau)  \, e^{-\frac{\pi}{\tau_2} \ep^2 m^2}.
\label{La cM explicit}
\end{align}
Then, $\La_{\cM[\ep]}$ should be 
finite for $\any \ep >0$.
Indeed, the $m=0$ term is absent due to the condition {\bf (1)}, and 
the potential UV-divergence around $\tau_2 \sim 0$ is removed 
by the damping factor $e^{-\frac{\pi}{\tau_2} \ep^2 m^2 }$  for $\any m \neq 0$,
as mentioned above.
Moreover, 
the convergence in the IR-region $\tau_2 \rightarrow +\infty$ is ensured by 
the condition {\bf (4)}.


Let us next prove the more non-trivial statement \eqn{La cM 0}. 
Set 
\begin{equation}
\bm{Z}_{(w,m)}(\tau; \ep) := Z_{(w,m)}(\tau) \, \frac{\ep}{\sqrt{\tau_2} \left|\eta(\tau)\right|^2}\, e^{- \frac{\pi}{\tau_2}\ep^2 \left|w\tau +m \right|^2} ,
\hspace{1cm} (\any w,m \in \bz).
\end{equation}
Because of the  modular covariance 
$$
\left. \bm{Z}_{(w,m)}(\tau; \ep)\right|_S = \bm{Z}_{(m,-w)}(\tau;\ep), \hspace{1cm}
\left. \bm{Z}_{(w,m)}(\tau; \ep)\right|_T = \bm{Z}_{(w,w+m)}(\tau;\ep),
$$
together with the condition {\bf (1)}, that is,  $\bm{Z}_{(0,0)}(\tau; \ep)\equiv 0$, 
we can rewrite  \eqn{La cM explicit} as
\begin{align}
\La_{\cM[\ep]} 
& \equiv \frac{1}{V_D} \int_{\cS} \frac{d^2\tau}{\tau^2_2}\,  \sum_{m\in\bz}\, \bm{Z}_{(0,m)}(\tau;\ep) 
= \frac{1}{V_D} \int_{\cF} \frac{d^2\tau}{\tau^2_2}\,  \sum_{w, m\in\bz}\, \bm{Z}_{(w,m)}(\tau;\ep) ,
\label{eval La ep 1}
\end{align}
due to the arguments given in 
\cite{Polchinski,OBrien,McClain},
which are commonly used in thermal string theory. 
Introducing the `Fourier transform', 
\begin{align}
\hZ_{(\al,\beta)}(\tau) := \frac{1}{N} 
\sum_{a,b\in \bz_N}\, Z_{(a,b)}(\tau) e^{2\pi i \frac{1}{N} \left(\al b - \beta a\right)},
\label{hZ}
\end{align}
and making use of the Poisson resummation, 
we can further rewrite the R.H.S of \eqn{eval La ep 1} as follows,
\begin{align}
\La_{\cM[\ep]} 
& = \frac{1}{V_D} \int_{\cF} \frac{d^2\tau}{\tau^2_2}\, \frac{1}{N \ep} \frac{1}{\sqrt{\tau_2} \left|\eta(\tau)\right|^2} \sum_{w, m\in\bz}\, 
\hZ_{(w,m)}(\tau) e^{-\frac{\pi}{\tau_2} \frac{1}{ N^2 \ep^2} \left| w\tau+ m\right|^2}
\nn
& = \frac{1}{V_D} \int_{\cF} \frac{d^2\tau}{\tau^2_2}\, \frac{1}{N \ep} \frac{1}{\sqrt{\tau_2} \left|\eta(\tau)\right|^2}
 \sum_{w, m\in\bz, \atop  (w,m)\neq (0,0)}\,
\hZ_{(w,m)}(\tau) e^{-\frac{\pi}{\tau_2} \frac{1}{ N^2 \ep^2} \left| w\tau+ m\right|^2}.
\label{eval La ep 2}
\end{align}
In the second line we made use of the fact that
$$
\hZ_{(0,0)}(\tau) \equiv \frac{1}{N}\sum_{a,b\in \bz_N}\, Z_{(a,b)}(\tau) = 0,
$$
due to the condition {\bf (5)}. 

We first estimate the non-zero winding sectors 
with
$w\neq 0$ in \eqn{eval La ep 2}.
For each fixed $w \neq 0$, the summation over $m\in \bz$ is evaluated by using the Poisson resummation, giving the inequality,
\begin{align}
& \frac{1}{V_D N \ep} \frac{1}{\sqrt{\tau_2} \left|\eta(\tau)\right|^2}\,
 \sum_{m\in\bz}\, \left|\hZ_{(w,m)}(\tau) \right| \, e^{-\frac{\pi}{\tau_2} \frac{1}{ N^2 \ep^2} \left| w\tau+ m\right|^2}
\leq \frac{\cC}{\ep} \, \frac{1}{\tau_2^{\frac{D-2}{2}}} \, e^{-4 \pi \tau_2 \left(h_0 - \frac{1}{2}\right)}\, e^{-\pi\frac{1}{ N^2 \ep^2} w^2 \tau_2}
\nn
&
\hspace{10cm} (\any w\neq 0, ~ \any \tau \in \cF) ,
\nonumber
\end{align}
with some finite constant $\cC$, $h_0$. 
Here, $h_0$ is the lowest conformal weight in the discrete part of the spectrum and {\em allowed to be tachyonic.}

Therefore, 
replacing the integration region $\cF$ with the slightly larger one,
\begin{equation}
\left\{ \tau \in \bh^+\, :\, -\frac{1}{2} \leq \tau_1 < \frac{1}{2}, ~ \tau_2 >  \frac{\sqrt{3}}{2} \right\},
\label{int region}
\end{equation}
we find 
\begin{align}
\left| \La_{\ep, w} \right| 
& \leq \frac{\cC}{\ep} \int_{\sqrt{3}/2}^{\infty} \frac{dt}{t^2} \frac{1}{t^{\frac{D-2}{2}}}\, 
 e^{-4 \pi t \left(h_0 - \frac{1}{2}\right)}\, e^{-\pi\frac{1}{ N^2 \ep^2} w^2 t}  
\hspace{1cm} (\any w\neq 0) .
\nonumber
\end{align}
This integral 
is obviously finite 
for sufficiently small $\ep >0$, and we readily obtain
\begin{equation}
\lim_{\ep\rightarrow +0} \, \sum_{w\neq 0}\, \La_{\ep, w} =0.
\label{limit La w}
\end{equation}


We next focus on the $w=0$ contribution.
Replacing again the integration region with \eqn{int region},
we obtain the following evaluation 
\begin{align}
\left| \La_{\ep, w=0} \right| & \equiv \frac{1}{N \ep V_D} \left| \int_{\cF} \frac{d^2\tau}{\tau^2_2}\, 
\frac{1}{\sqrt{\tau_2} \left|\eta(\tau)\right|^2}\,
\sum_{m\neq 0} \, \hZ_{(0,m)}(\tau)  
e^{-\frac{\pi}{\tau_2} \frac{m^2}{N^2 \ep^2}} \right|
\nn
& \leq \frac{\cC'}{\ep} 
\int_{\sqrt{3}/2}^{\infty} \frac{dt}{t^{\frac{5}{2}}} \,  \frac{1}{\left|\eta(it)\right|^2}\, 
\left| \sum_{m\neq 0} \,  \left. 
\hZ_{(0,m)}(it)\right|_{\msc{level-matched}}\right|  \, 
e^{-\frac{\pi}{t} \frac{m^2}{N^2 \ep^2}},
\label{eval La ep 3}
\end{align} 
with some finite constant $\cC'$.
Moreover, we note that 
$$
\left. e^{4\pi \frac{1}{24} \tau_2} \, \hZ_{(0,m)}(\tau)\right|_{\msc{level-matched}} 
\equiv e^{4\pi \frac{1}{24} \tau_2} \,  \frac{1}{N} \sum_{a,b\in \bz_N}\, \left.
Z_{(a,b)}(\tau) \right|_{\msc{level-matched}}
 e^{-2\pi i \frac{m}{N}a},
$$
is non-tachyonic due to the condition {\bf (4)}.
Thus, 
we obtain the following evaluation with a finite constant $\cC''$;
\begin{align}
\left| \La_{\ep, w=0} \right| & \leq  \frac{\cC''}{\ep} \int_{\sqrt{3}/2}^{\infty} \frac{dt}{t^{\frac{5}{2}}} \, \frac{1}{t^{\frac{D-2}{2}}}\,
  \sum_{m=1}^{\infty}\, 
e^{-\frac{\pi}{t} \frac{m^2}{N^2 \ep^2}}
\nn
& < \frac{\cC''}{\ep} \pi^{- \frac{D+1}{2}} \left(N\ep \right)^{D+1}\,
\int_0^{\infty} \frac{ds}{s} \, s^{\frac{D+1}{2}} \, \sum_{m=1}^{\infty} \, e^{-m^2 s}\, 
\nn
& 
= \cC'' N^{D+1} \ep^{D} \, \widehat{\zeta}(D+1),
\end{align}
where $\widehat{\zeta} (s) \equiv \pi^{-\frac{s}{2}} \Gamma\left(\frac{s}{2}\right) \zeta(s)$ is the completed zeta function. 

In this way, as long as $D \geq 1$, we obtain the desired result;
$$
\lim_{\ep \, \rightarrow\, +0} \, \La_{\cM[\ep]} \equiv  \lim_{\ep \, \rightarrow\, +0} \, \left[\La_{\ep, w=0} + \sum_{w\neq 0} \La_{\ep, w} \right] =0. 
\hspace{2cm}
\mbox{\bf (Q.E.D)}
$$

~


A few comments are in order:
\begin{description}
\item[(i)] 
Under the requirement {\bf (2)}
the bosonic and fermionic mass spectra in the point particle theory $\cM[\ep]$
are obviously mismatched. 
Nevertheless, the 1-loop cosmological constant $\La_{\cM[\ep]}$ vanishes 
under the $\ep\, \rightarrow \, +0$ limit in the cases of $D \geq 1$.


\item[(ii)] 
As $\ep$ plays the role of the UV cut-off, 
the integration region is truncated roughly as follows,
$$
\int_{\cS} \frac{d^2\tau}{\tau_2^2} ~ \longrightarrow ~ \int_{\ep^2}^{\infty} \frac{d\tau_2}{\tau_2^2} \, \int_{-1/2}^{1/2} d\tau_1.
$$
In other words, we have a {\em substringy\/} cut-off mass scale 
$
\dsp
M_{\ep} \sim \frac{1}{\ep \sqrt{\al'}} \left( \equiv \frac{1}{\ep} M_s \right),
$
and the low energy mass spectrum $M \ll M_{\ep}$ is 
nearly equal to that of $\dsp Z_0(\tau) \equiv \frac{1}{N} \sum_{m\in \bz_N}\, Z_{(0,m)}(\tau) $, 
which only contains the $g$-invariant particle spectrum. 
On the other hand, 
the `super-partners'  that are not $g$-invariant acquire
the masses greater than $M_{\ep}$.
Consequently, $M_{\ep}$ is  interpreted as the energy scale of the spontaneously broken SUSY
in our theory $\cM[\ep]$.
It is remarkable that, 
{\em while the naive dimensional estimation implies 
$$
\La_{\cM[\ep]} \sim M_{\ep}^D \equiv \ep^{-D} M_s^D,
$$
we indeed obtain 
$$
\La_{\cM[\ep]} \sim \ep^D M_s^D \equiv \ep^{2D} M_{\ep}^D, $$ 
}
as was shown above. 
The existence of the suppression  factor $\ep^{2D}$ is crucial in our models. 

\item[(iii)] 
Intuitively, the counterpart of the bose-fermi cancellation in the original 
orbifold
model is disguised by the modular transformation 
as other particles 
in the resultant 
particle theory $\cM[\ep]$, which contribute to $\La_{\cM[\ep]} $ in the integration 
region $\cS$ outside $\cF$.

\end{description}

~


\section{Simple Models from String Vacua}
\label{sec:simple models}

In this section we exhibit  simple examples 
satisfying the requirements {\bf (1)} $\sim$ {\bf (5)} given in the previous section 
based on the toroidal asymmetric orbifolds of type II or heterotic superstring vacua.  
To this end we shall make use of the modular invariants, or modular covariant functions more generally,
constructed in \cite{SSLie}. 
They are associated to  Lie algebra lattices satisfying the `self-duality condition'  under the T-duality twist. 
We start with reviewing  briefly the results  given in \cite{SSLie}
for our construction of the relevant models.
Since we only need appropriate modular forms, 
slight extensions and refinements are also included below.

~


\subsection{Asymmetric Orbifolds based on Lie Algebra Lattices}
\label{review asymm}

Let us consider the $r$-dimensional torus $T^r[X_r]$ 
associated to the Englert-Neveu
lattice \cite{EN} (see also \cite{Lerche lattice} for a review)
for  the semi-simple Lie algebra 
$X_r$ (rank $r$), for which we have the symmetry enhancement to the affine $X_r$-symmetry with level 1 on the string world-sheet. 
We simply call it the `Lie algebra lattice for $X_r$' in this section. 
We assume that the simple parts of $X_r$ are composed only of 
\begin{equation}
A_1, ~ E_7, ~ D_n~ (\any n \in \bz_{>0}) .  
\label{AED}
\end{equation}
Here, we have slightly extended the argument  in \cite{SSLie}
by  including $D_n$
with odd $n$.
The number of `$D_{\msc{odd}}$-components' 
is supposed to be even to preserve the unitarity, as explained below.  
Then the asymmetric orbifold model of $T^r[X_r]$ for the chiral reflection (or the `T-duality transformation')
$$
(\-_R)^{\otimes r} ~ : ~ \left(X^i_L, X^i_R\right) ~ \longrightarrow ~  \left(X^i_L, - X^i_R\right) 
$$
is described by  the following partition function
\cite{SSLie},
\begin{align}
& Z^{T^r[X_r]}_{\msc{orb}} (\tau) \equiv \frac{1}{16} \sum_{a,b \in \bz_{16}} \, Z^{T^r[X_r]}_{(a,b)} (\tau),
\label{Z TXr orb}
\end{align}
where
\begin{align}
 Z^{T^r[X_r]}_{(a,b)} (\tau) &  \equiv
\left\{
\begin{array}{ll}
\dsp
 \ep^{[r]}_{(a,b)} \, \overline{\left( \tchi^{A_1}_{(a,b)} (\tau) \right)^r} \,
\chi^{X_r}_{(a,b)} (\tau) 
& ~~
\left(a \in 2\bz+1 ~ \mbox{or} ~ b \in 2\bz+1 \right) ,
\\
\dsp 
Z^{T^r[X_r]} (\tau)  
& ~~ \left(a, b \in 2\bz\right) ,
\end{array}
\right.
\label{Z TXr ab}
\end{align} 
and we introduced the notation
\begin{align}
\tchi^{A_1}_{(a,b)}(\tau) & := 
\left\{
\begin{array}{ll}
\sqrt{
\frac{\th_3 \th_4}{\eta^2}
}
& ~~ (a\in 2\bz, ~ b\in 2\bz+1) , \\
\sqrt{
\frac{\th_3 \th_2}{\eta^2}
}
& ~~ (a\in 2\bz+1, ~ b\in 2\bz) , \\
\sqrt{
\frac{\th_4 \th_2}{\eta^2}
}
& ~~ (a\in 2\bz+1, ~ b\in 2\bz+1) , \\
\end{array}
\right.
\label{tchi A1ab}
\\
\ep^{[r]}_{(a,b)} & := \left\{
\begin{array}{ll}
e^{\frac{i \pi r}{8}ab}
& ~~ \left(a \in 2\bz, ~ b \in 2\bz+1\right) ,
\\
e^{-\frac{i \pi r}{8}ab}
& ~~ \left(a \in 2\bz+1, ~ b \in 2\bz\right) ,
\\
\left(\kappa_a\right)^r 
e^{-\frac{i \pi r}{8}ab}
& ~~ \left(a , ~ b \in 2\bz+1\right) ,
\end{array}
\right.
\label{ep r ab}
\\
\kappa_a & :=
e^{-\frac{i\pi }{8}(a^2-1 )} 
\equiv
 \left\{
\begin{array}{ll}
+1 & ~~ (a \equiv 1,7 ~ \mod \, 8) ,
\\
-1 & ~~ (a \equiv 3,5 ~ \mod \,  8) .
\end{array}
\right.
\end{align}
Moreover, $\chi^{X_r}_{(a,b)}(\tau)$ is defined as the simple product 
$$
\chi^{X_r}_{(a,b)}(\tau) := \prod_{i} \, \chi^{X_{r_i}^{(i)}}_{(a,b)} (\tau) , 
\hspace{1cm} \sum_i \, r_i = r, 
$$
and each component is
associated to the simple Lie algebra $X_{r_i}^{(i)}$ appearing in \eqn{AED}.
The explicit forms of  $Z^{T^r[X_r]} (\tau)$
and  $\chi^{X_r}_{(a,b)}(\tau)$ are summarized in appendix A.


One can derive \eqn{Z TXr ab} by first evaluating the trace,
$$
Z^{T^r[X_r]}_{(0,1)}(\tau) \equiv \tr \left[\, \left(\-_R\right)^{\otimes r} q^{L_0-\frac{c}{24}} \overline{q^{\tL_0-\frac{c}{24}}}\, \right] ,
$$
with $c=r$,
and by requiring the modular covariance,
\begin{equation}
\left. Z^{T^r[X_r]}_{(a,b)}(\tau)\right|_T = Z^{T^r[X_r]}_{(a,a+b)}(\tau),
\hspace{1cm}
\left. Z^{T^r[X_r]}_{(a,b)}(\tau)\right|_S = Z^{T^r[X_r]}_{(b,-a)}(\tau),
\label{mod cov ZTXr}
\end{equation}
which ensures the modular invariance of \eqn{Z TXr orb}.
See \cite{SSLie} for details. 
We emphasize that  the non-trivial phase factor $\ep^{[r]}_{(a,b)}$  is necessary to achieve 
the  modular covariance \eqn{mod cov ZTXr},
and  thus, the building blocks  \eqn{Z TXr ab} generically possess an order 16 periodicity, 
rather than the naive expectation of
order 2, for the twisted sectors with $a\neq 0$.
These phases appear because  the action of the 
chiral 
reflection in the target space is uplifted 
 on the world-sheet
(see e.g.\cite{Aoki:2004sm,SSW,Harvey:2017rko}),
and will play a crucial role in our argument given below.

We also note another interpretation of the asymmetric orbifolds given above. 
The Narain lattice defining the torus $T^r[X_r]$ can be generically decomposed as 
\begin{align}
\Gamma^{r,r}[X_r] & = \sum_{\al_{L}, \, \al^{(i)}_{R}}
\La_{(\al_L)}^{X_r} \bigoplus \sum_{i=1}^r \, \La^{A_1}_{(\al_R^{(i)})}
\equiv \La_{(0)}^{X_r} \bigoplus \left[ \La^{A_1}_{(0)} \oplus \cdots \oplus \La^{A_1}_{(0)} \right] + \cdots ,
\label{lattice decomp 1}
\end{align}
where `$\La^{X_r}_{(\al)}$' denotes the Lie algebra lattice of $X_r$ associated to the conjugacy class labeled by 
$\al$ (corresponding to a certain integrable representation
 of affine $X_r$ with level 1).
Especially, $\La^{X_r}_{(0)}$ is nothing but the root lattice of $X_r$ associated to the basic representation. 
Only  the term including  $\left[ \La^{A_1}_{(0)} \oplus \cdots \oplus \La^{A_1}_{(0)}\right]$
can contribute to the trace $\dsp \tr \left[\left(\-_R\right)^{\otimes r} \, \cdots \right]$ for the untwisted sector
in this decomposition.
If focusing on each of the right-moving momentum lattice $\La^{A_1}_{(\al)}$, the chiral reflection 
$\-_R$ is naturally described by the $SU(2)$-current algebra $\{ J_R^a \}$ of level 1.
Namely, one can simply identify 
$\dsp \-_R \equiv e^{i\pi J^1_{R,0}}$ 
in the basic
representation (see appendix B).
One can also adopt the `chiral half-shift' $s_R \equiv e^{i\pi J^3_{R,0}}$ as the involutive operator  
in the untwisted sector
to define the relevant 
asymmetric orbifolds, 
and obtains the same  building blocks $Z^{T^r[X_r]}_{(a,b)}(\tau)$
because of the obvious reason of the $SU(2)$ invariance\footnote
   {More precisely, the chiral half-shift operator $s_R$ (as well as the chiral reflection) has to contain the phase factor 
to achieve the involutive property $s_R^2 = \bm{1}$.
Namely, we should adopt the definition 
$
s_R \equiv e^{-\frac{i\pi}{2} \ell}\, e^{i\pi J^3_{R, 0}},
$ 
for the spin $\ell/2$ representation of $\widehat{A_1}$ with level 1 ($\ell=0,1$).
}.

In the  following,  we will mainly focus on the chiral half-shifts rather than the chiral reflections
to obtain desired models, which modifies the construction in \cite{SSLie}. 
We will also maintain the world-sheet $\cN=1$ superconformal symmetry,
since we expect that the consistency of
the original superstring theories is important to ensure 
that of the resultant particle theories after interactions are turned on%
\footnote
{However, it may not be necessary to assume the world-sheet superconformal symmetry, since our purpose in this paper 
    is to construct the non-SUSY {\em particle theories\/}  with vanishing cosmological constant, rather than the non-SUSY string vacua. 
   In other words one may be able to relax the requirements of the consistency as string vacua.   
 We will again discuss this point in section 4.}.
The chiral half-shift preserves the superconformal symmetry 
even if trivially acting on the world-sheet fermions 
$\psi^i_R$,
whereas we need to require
$$
\left(\-_R\right)^{\otimes r} ~: ~ (\psi_L^i, \psi_R^i) ~ \longrightarrow ~ (\psi_L^i, - \psi_R^i),
$$
for the chiral reflection.

~



\subsection{Type II Models}

Let us start with the type II string on $\br^{3,1} \times T^5 \times S^1_{N \ep}$, where $N$ is the order of the $g$-orbifold introduced below. 
In other words we focus on the cases with $\cK = T^5$ in the notations given in section \ref{sec 2}. 
As discussed in section \ref{sec 2}, 
we consider the orbifold  by $\bm{g} \equiv g \otimes \cT_{2\pi \ep}$ with 
appropriate choices of $g$ acting on the $T^5$-sector as well as the world-sheet fermions. 
The modular invariant partition function of the type II string on this orbifold is written in the form as 
\eqn{total ZN-orb}, and we shall only focus  on the building block `$Z_{(a,b)}(\tau)$' in \eqn{total ZN-orb},
which includes the contributions from  $\br^{3,1} \times T^5$
with the  $g^a$($g^b$)-twist along the spatial(temporal) direction.

We consider the orbifolding by the following involutive operator 
with $N_0=2$ in the notation of section \ref{sec 2},  
\begin{equation}
g:=  (-1)^{F_L+F_R}  \otimes s_R[k+r] \otimes s_L[k+\ell],
\label{g type II}
\end{equation}
where {\em e.g.} `$s_R[m]$' denotes the chiral half-shift along the $m$ directions in $T^5$ and we set
$d+k+\ell+r =5$ $(d,k,\ell,r \geq 0)$. 
We  consider the case where  
the Narain lattice of $T^5$ can be decomposed as 
\begin{align}
\Gamma^{5,5} & = \Gamma^{d,d} \bigoplus 
\left[\left(\La^{A_1}_{(0)}\right)^{k+\ell} \oplus \La^{X_{r}}_{(0)} \right] \bigoplus 
\left[\left(\La^{A_1}_{(0)}\right)^{k+r} \oplus \La^{X'_{\ell}}_{(0)} \right] 
+ \cdots ,
\label{decomp Gamma 66}
\end{align}
with some Lie algebra lattices for $X_r$, $X'_{\ell}$ 
with rank $r$, $\ell$
in order to make the operator $g$ well-defined. 
The standard symbol $(-1)^{F_L}$ ($(-1)^{F_R}$) acts on the left(right)-moving  Ramond sector as  the sign-flip  (`space-time fermion number mod 2').
We assume that $X_r$, $X'_{\ell}$ are composed only of $A_1$, 
$D_n$, and 
 the total number of the $D_n$  components with $n \in 2\bz+1$
 in  $X_r$, $X'_{\ell}$
 should be even to maintain unitarity condition {\bf (2)}, as explained shortly. 

The building block $Z_{(a,b)}(\tau)$ is explicitly written in terms of the twisted characters given in  
section \ref{review asymm} (see also appendix A).
For the  `even sectors', {\em i.e.} sectors with $a,b\in 2\bz$, 
the building blocks
are equal to the partition function of 
the  unorbifolded model. 
On the other hand, 
those for 
the `odd sectors',  {\em i.e.} sectors with $a\in 2\bz+1$ or $b \in 2\bz+1$,
are non-trivial. They are explicitly written as 
\begin{align}
Z_{(a,b)}(\tau) & = 
\frac{1}{\tau_2 \left|\eta(\tau)\right|^4}\, 
Z^{T^d}(\tau) \cdot \left|h_{(a,b)}\right|^2 \cdot
\left|\tchi^{A_1}_{(a,b)}\right|^{2k} 
\nn
& \hspace{1cm}
\times  \ep^{[r]}_{(a,b)} \, 
\overline{\left(\tchi^{A_1}_{(a,b)}\right)^{r}} \, \chi^{X_r}_{(a,b)} \cdot
\ep^{[-\ell]}_{(a,b)} \, 
\left(\tchi^{A_1}_{(a,b)}\right)^{\ell} \, \overline{\chi^{X'_{\ell}}_{(a,b)}} 
\nn
& = \frac{1}{\tau_2 \left|\eta(\tau)\right|^4}\, Z^{T^d}(\tau) 
\cdot 
\left|h_{(a,b)}\right|^2
\nn
& \hspace{1cm} \times
\ep^{[r-\ell]}_{(a,b)} \,
\overline{\left(\tchi^{A_1}_{(a,b)}\right)^{r+k}} \,
\left(\tchi^{A_1}_{(a,b)}\right)^{\ell+k} \, 
\chi^{X_r}_{(a,b)} \, \overline{\chi^{X'_{\ell}}_{(a,b)}} .
\label{type II ab general}
\end{align}
Here we introduced the free fermion chiral blocks twisted by $(-1)^{F_L}$, 
\begin{align}
h_{(a,b)} & := \left\{
\begin{array}{ll}
\left(\frac{\th_3}{\eta}\right)^4 -  \left(\frac{\th_4}{\eta}\right)^4 + \left(\frac{\th_2}{\eta}\right)^4
& ~~ \left(a \in 2\bz, ~ b \in 2\bz+1\right) ,
\\
 \left(\frac{\th_3}{\eta}\right)^4 +  \left(\frac{\th_4}{\eta}\right)^4 - \left(\frac{\th_2}{\eta}\right)^4
& ~~ \left(a \in 2\bz+1, ~ b \in 2\bz\right) ,
\\
 - \left[ \left(\frac{\th_3}{\eta}\right)^4 +  \left(\frac{\th_4}{\eta}\right)^4 + \left(\frac{\th_2}{\eta}\right)^4 \right]
& ~~ \left(a , ~ b \in 2\bz+1\right) .
\end{array}
\right.
\end{align}

One can readily confirm the modular covariance,
\begin{align}
\left. Z_{(a,b)}(\tau)\right|_{T} = Z_{(a,a+b)}(\tau),
\hspace{1cm}
\left. Z_{(a,b)}(\tau)\right|_{S} = Z_{(b,-a)}(\tau),
\label{mod cov type II}
\end{align}
which ensures the modular invariance of the total partition function. 
In fact, the first line on
the R.H.S of \eqn{type II ab general} is written in the form that is manifestly modular covariant. 
We again emphasize the importance of
the phase factor $\ep^{[*]}_{(a,b)}$ to achieve this relation. 

~


Now, we discuss whether   
the model obtained from this setup
satisfies 
the requirements {\bf (1)} $\sim$ {\bf (5)} given in the previous section.
First of all, it is easy to check that {\bf (1)} $\sim$ {\bf (3)} are satisfied. 
We note that $g^2 = {\bf 1}$ in the untwisted sector, and the operator $(-1)^{F_L+F_R}$ 
removes, say, all the gravitinos in the spectrum read off from $\dsp Z_0(\tau) \equiv \frac{1}{2} \sum_{b\in \bz_2}\, 
Z_{(0,b)}(\tau)$. We also note that $\ep^{[*]}_{(0,b)}=1$ \footnote
   {The assumption that `the number of $D_n$-components with $n\in 2\bz+1$ should be even' was necessary for this statement.}, thus $Z_0(\tau)$ clearly yields a unitary 
spectrum.

We next focus on the more non-trivial requirement {\bf (5)}. 
For the `even sectors' with $a,b \in 2\bz$, 
the building blocks $Z_{(a,b)}(\tau)$
vanish as they
are the same as the partition function of the original supersymmetric 
model. 
We thus need to show that the summation over the `odd sectors'  
with $a \in 2\bz+1$ or $b \in 2\bz+1$ vanishes;
\begin{equation}
\sum_{\msc{$a \in 2\bz+1$ or $b\in 2\bz+1$}}\,  Z_{(a,b)} (\tau) =0.
\label{sum odd sec}
\end{equation} 
To do so it is sufficient to confirm the  following `criterion',
\begin{equation}
\sum_{b\in 2\bz \cap \bz_{16}}\, Z_{(a,b)} (\tau) =0 
\hspace{1cm} 
\left(\any a \in 2\bz+1\right) .
\label{criterion}
\end{equation}
Indeed, if this is the case, we readily obtain \eqn{sum odd sec}
due to the modular covariance \eqn{mod cov type II}
\footnote{
Almost the same argument has been  used in \cite{AS,AS2} in a different context. 
}.


Therefore, the question is what type of the
Lie algebra lattices for $X_r$, $X'_{\ell}$ 
in \eqn{type II ab general}
can satisfy the criterion \eqn{criterion}. 
We note that 
each term of the theta functions 
appearing in 
$Z_{(a,b)} (\tau)$
contains the various phase factors of the form
$
e^{\frac{i\pi}{8} K a b}
$
with $K \equiv r- \ell ~ (\mod \, 2)$.
(Recall the definitions of phase $\ep^{[*]}_{(a,b)}$ and the twisted characters  
$\tchi^{A_1}_{(a,b)}$, $\chi^{X_r}_{(a,b)}$ given in appendix A.
)
Thus, 
it is enough that all the theta function terms contain a {\em non-vanishing\/} phase factor in order
to satisfy \eqn{criterion}. 
Although it seems hard to fully classify the general choices of $X_r$, $X'_{\ell}$ satisfying \eqn{criterion}, 
the following two simple cases are obviously sufficient\footnote
   {However, we would like to emphasize that these two are sufficient but not necessary. 
Namely, there are many examples 
 that satisfy \eqn{criterion} apart from these cases {\bf (i)} and {\bf (ii)}.
};
\begin{description}
\item[(i)]
$\ell=0$, $r \not\in 8\bz $ ~($r=0$, $\ell \not\in 8\bz)$, and  
$X_r$ ($X'_{\ell}$) is simple ({\em i.e.} made up only of 
a single piece of $A_1$ (when $r=1$) or $D_r$).

\item[(ii)]
$r-\ell \in 2\bz+1$.

\end{description}


Finally, let us examine the requirement {\bf (4)}. 
At first glance, it seems that the tachyonic behavior would appear due to the fermion chiral block $h_{(a,b)}$ with $a \in 2\bz+1$. 
However, 
once \eqn{criterion} is satisfied, 
by using its T-transformation 
 we eventually obtain
$$
\sum_{b\in \bz_{16}}\, Z_{(a,b)} (\tau) = 0.
$$
Moreover, it is easy to show that we do not have any tachyonic modes in all the sectors of $a\in 2\bz$ 
from \eqn{criterion}.

~


\subsection{Heterotic Models}

We next try to construct the models
based on the heterotic string 
compactified on $T^5 \times S^1_{N\ep}$, where the left-mover is given by the 26 dimensional bosonic
theory.
We shall only consider the $E_8\times E_8$-cases, and the $SO(32)$-cases can be similarly treated.


We take the orbifold action 
\begin{align}
g & :=  (-1)^{F_R}  \otimes s_R[k+r] \otimes 
s_L[k+\ell_1+\ell_2+\ell_3]
\nn
& \equiv (-1)^{F_R}  \otimes s_R[k+r] \otimes s_L[k+\ell_1] \otimes s_L[\ell_2] \otimes s_L[\ell_3] ,
\label{g hetero}
\end{align}
where all the  notation is  defined as in \eqn{g type II}.
Here, we assume 
$d+k+\ell_1+r =5$, $\ell_2, \ell_3 \leq 8$.
$s_R[k+r]$, $s_L[k+\ell_1]$ act along the $T^5$-directions, while 
$s_L[\ell_2]$, $s_L[\ell_3]$ are assigned to the two $E_8$-factors. 
We consider the case where the total Narain lattice is
decomposed as 
\begin{align}
\Gamma^{21,5}  = & \Gamma^{d,d} \bigoplus 
\left[\left(\La^{A_1}_{(0)}\right)^{k+\ell_1+\ell_2+\ell_3} \oplus \La^{X_{r}}_{(0)} \oplus 
\La_{(0)}^{X_{8-\ell_2}} \oplus  
\La_{(0)}^{X_{8-\ell_3}}\right]
\nn
& 
 \bigoplus 
\left[\left(\La^{A_1}_{(0)}\right)^{k+r} \oplus \La^{X'_{\ell_1}}_{(0)} \right] 
+ \cdots .
\label{decomp Gamma 22 6}
\end{align}
Here,
$X_r$, $X'_{\ell_1}$ are composed only of $A_1$, $D_n$, and
$X_{8-\ell_2}$, $X_{8-\ell_3}$ can contain the pieces of 
$A_1$, $D_n$, $E_7$, or $E_8$.
The chiral half-shifts $s_L[*]$, $s_R[*]$ are associated to the $A_1$-pieces $\left(\La_{(0)}^{A_1}\right)^*$.
We again require that  
the number of total $D_n$ components with $n \in 2\bz+1$ 
in the left- and right-movers
should be even 
in order to satisfy the `unitarity condition' {\bf (3)}. 

The building blocks for the even sectors 
are again equal to the partition function of the unorbifolded model,
and those of the odd sectors are evaluated as 
\begin{align}
Z_{(a,b)}(\tau) & = 
\frac{1}{\tau_2 \left|\eta(\tau)\right|^4}\, 
Z^{T^d}(\tau) 
\cdot \overline{h_{(a,b)}} \cdot
\left|\tchi^{A_1}_{(a,b)}\right|^{2k} 
\nn
& \hspace{1cm}
\times  \ep^{[r]}_{(a,b)} \, 
\overline{\left(\tchi^{A_1}_{(a,b)}\right)^{r}} \, \chi^{X_r}_{(a,b)} \cdot
\ep^{[-\ell_1]}_{(a,b)} \, 
\left(\tchi^{A_1}_{(a,b)}\right)^{\ell_1} \, \overline{\chi^{X'_{\ell_1}}_{(a,b)}} 
\nn
& \hspace{1cm}
\times \prod_{i=2,3}\, \left[\ep^{[8-\ell_i]}_{(a,b)} \, (-1)^{ab}\, \left(\tchi_{(a,b)}^{A_1} \right)^{\ell_i} \, \chi^{X_{8-\ell_i}}_{(a,b)} \right]
\nn
& = \frac{1}{\tau_2 \left|\eta(\tau)\right|^4}\, Z^{T^d}(\tau) 
\cdot 
 \overline{h_{(a,b)}}
\nn
& \hspace{1cm} \times
\ep^{[r-\sum_{i} \ell_i]}_{(a,b)} \,
\overline{\left(\tchi^{A_1}_{(a,b)}\right)^{k+r}} \,
\left(\tchi^{A_1}_{(a,b)}\right)^{k+\sum_i \ell_i} \, 
\chi^{X_r}_{(a,b)} \, \overline{\chi^{X'_{\ell_1}}_{(a,b)}} \,
\prod_{i=2,3}\, \chi^{X_{8-\ell_i}}_{(a,b)} .
\label{hetero ab general}
\end{align}
The first line of \eqn{hetero ab general} shows its modular covariance manifestly\footnote
   {Note that $(-1)^{ab} \left[\tchi^{A_1}_{(a,b)}(\tau)\right]^8$ behaves  modular covariantly upto the phase factor
   arising from the $T$-transformation of $\eta(\tau)^{-8}$. }.
One can likewise show that the total partition function of this model generally satisfies {\bf (1)} $\sim$ {\bf (3)},
and the requirements {\bf (4)}, {\bf (5)} 
are achieved if the condition \eqn{criterion} is satisfied. 
As in the type II models, we can exhibit the two simple cases satisfying \eqn{criterion}, though the general classification 
would be hard to describe;
\begin{description}

\item[(i)]
Only one of $\{ r, \ell_1, \ell_2, \ell_3\}$ is non-zero (mod 8), and the Lie algebra lattice corresponding to the non-vanishing integer 
({\em e.g.}  $X_{8-\ell_2}$ when $\ell_2 \not\equiv 0 ~ (\mod \, 8)$)  is simple ({\em i.e.} made up only of 
a single piece of $A_1$, $D_r$ or $E_7$).

\item[(ii)]
$\dsp r-\sum_{i=1}^3 \ell_i \in 2\bz+1$.

\end{description}


~


\subsection{Simple Examples}
\label{subsec: ex}

Here we present simple examples to demonstrate the general features given in the previous subsections.

~


\noindent
{\bf 1. Example from type II string on
$T^4[D_4] \times S^1 \times S^1_{N\ep}$ : }

~

The first example we consider is 
a particle model from
the type II string on 
$
\br^{3,1} \times   T^4[D_4] \times S^1 \times S^1_{N\ep},
$
where 
$T^4[D_4]$ is the 4-dim. torus for
the Englert-Neveu lattice of $D_4$, in other words, 
at the $SO(8)$-symmetry enhancement point. 
$S^1$ is a circle with an arbitrary radius, 
which is not important below.  
We simply choose $k=\ell=0$, $r=4$, that is, the orbifold action is defined as 
\begin{equation}
g= (-1)^{F_L+F_R} \otimes s_R[4],
\label{g ex1}
\end{equation}
where $s_R[4]$ acts on $T^4[D_4]$.

The building blocks \eqn{type II ab general} for the odd sectors 
with  $a \in 2\bz+1$ or $b\in 2\bz+1$ are written in this case as 
\begin{align}
Z_{(a,b)}(\tau) & = 
\frac{1}{\tau_2 \left|\eta(\tau)\right|^4}\, 
Z^{S^1}(\tau) \cdot \left|h_{(a,b)}\right|^2 
\cdot
  \ep^{[4]}_{(a,b)} \, 
\overline{\left(\tchi^{A_1}_{(a,b)}\right)^{4}} \, \chi^{D_4}_{(a,b)} .
\label{ex1 ab general}
\end{align}
Especially, for the sectors with $a \in 2\bz+1$, $b\in 2\bz$,
we can explicitly write it down as  
\begin{align}
Z_{(a,b)}(\tau) & = 
\frac{1}{\tau_2 \left|\eta(\tau)\right|^4}\, 
Z^{S^1}(\tau) \cdot 
\left|
\left(\frac{\th_3}{\eta}\right)^4 + \left(\frac{\th_4}{\eta}\right)^4 -\left(\frac{\th_2}{\eta}\right)^4
\right|^2
\nn
& \hspace{1cm} \times  (-1)^{\frac{b}{2}}\, 
\overline{\left(\frac{\th_3 \th_2}{\eta^2}\right)^2} \, 
\frac{1}{2} \left[\left(\frac{\th_3}{\eta}\right)^4 + \left(\frac{\th_2}{\eta}\right)^4\right] .
\label{ex1 ab odd}
\end{align}
Due to the existence of 
the phase factor $(-1)^\frac{b}{2}$ the criterion \eqn{criterion} is obviously satisfied;
$$
\sum_{b\,:\,\msc{even}}\, Z_{(a,b)}(\tau) =0.
$$

On the other hand, the partition function of the untwisted sector ($a=0$) is evaluated as 
\begin{align}
\hspace{-0.5cm}
Z_0(\tau)
& = \frac{1}{2}
\frac{1}{\tau_2 \left|\eta(\tau)\right|^4}\, 
Z^{S^1}(\tau) \cdot 
\left|
\left(\frac{\th_3}{\eta}\right)^4 - \left(\frac{\th_4}{\eta}\right)^4 - \left(\frac{\th_2}{\eta}\right)^4
\right|^2
\frac{1}{2} \left[ \left|\frac{\th_3}{\eta}\right|^8 
+ \left|\frac{\th_4}{\eta}\right|^8
+ \left|\frac{\th_2}{\eta}\right|^8
\right]
\nn
& + \frac{1}{2}
\frac{1}{\tau_2 \left|\eta(\tau)\right|^4}\, 
Z^{S^1}(\tau) \cdot 
\left|
\left(\frac{\th_3}{\eta}\right)^4 - \left(\frac{\th_4}{\eta}\right)^4 +\left(\frac{\th_2}{\eta}\right)^4
\right|^2
\overline{\left(\frac{\th_3 \th_4}{\eta^2}\right)^2} \, 
\frac{1}{2} \left[\left(\frac{\th_3}{\eta}\right)^4 + \left(\frac{\th_4}{\eta}\right)^4\right] ,
\nn
& 
\label{ex1 a=0 b}
\end{align}
which does not identically vanish. 
In fact, 
this partition function describes 
128 massless space-time bosons in the standard fashion, while all the space-time fermions get massive. 
Especially, all of the space-time fermions belonging to the massless super-multiplets 
in the unorbifolded theory are removed by the $g$-projection, which means that they 
acquire the masses of order $\dsp M_{\ep} \sim 
\frac{1}{\ep \sqrt{\al'}} $ in the theory $\cM[\ep]$. 
(See the comment {\bf (ii)} in section 2.)
Thus, the bose-fermi cancellation is `maximally' broken in 
$\cM[\ep]$.
Nevertheless, the cosmological constant $\La_{\cM[\ep]}$ vanishes under the $\ep\, \rightarrow\, +0$ limit, as we have proven in section 2.


~


\noindent
{\bf 2. Counter example from type II string on
$ T^4[D_2 \oplus D_2] \times S^1 \times S^1_{N\ep}  $}

~

As a digression, here we give an example which does not
satisfy our criterion. 
We adopt almost the same background as the first one, but with the $T^4[D_4]$ replaced with 
\begin{equation}
T^4[D_2 \oplus D_2] \left( \cong T^4 [A_1 \oplus A_1 \oplus A_1 \oplus A_1] \right),
\end{equation}
and the orbifold action is again given by \eqn{g ex1}.

The total building blocks \eqn{type II ab general} now become
\begin{align}
Z_{(a,b)}(\tau) & = 
\frac{1}{\tau_2 \left|\eta(\tau)\right|^4}\, 
Z^{S^1}(\tau) \cdot \left|h_{(a,b)}\right|^2 
\cdot
  \ep^{[4]}_{(a,b)} \, 
\overline{\left(\tchi^{A_1}_{(a,b)}\right)^{4}} \, \left[\chi^{D_2}_{(a,b)} \right]^2,
\label{ex2 ab general}
\end{align}
and  we  obtain for the sectors with $a \in 2\bz+1$, $b\in 2\bz$,
\begin{align}
Z_{(a,b)}(\tau) & = 
\frac{1}{\tau_2 \left|\eta(\tau)\right|^4}\, 
Z^{S^1}(\tau) \cdot 
\left|
\left(\frac{\th_3}{\eta}\right)^4 + \left(\frac{\th_4}{\eta}\right)^4 -\left(\frac{\th_2}{\eta}\right)^4
\right|^2
\nn
& \hspace{1cm} \times  (-1)^{\frac{b}{2}}\, 
\overline{\left(\frac{\th_3 \th_2}{\eta^2}\right)^2} \, 
\frac{1}{4} \left[\left(\frac{\th_3}{\eta}\right)^4 + \left(\frac{\th_{2}}{\eta}\right)^4 
+ 2 (-1)^{\frac{b}{2}} \left(\frac{\th_3 \th_2}{\eta^2}\right)^2\right] ,
\label{ex2 ab odd}
\end{align}
in place of \eqn{ex1 ab odd}.
We thus find that
$$
\sum_{b\,:\,\msc{even}}\, Z_{(a,b)}(\tau)
\propto 
\frac{1}{\tau_2 \left|\eta(\tau)\right|^4}\, 
Z^{S^1}(\tau) \cdot 
\left|
\left(\frac{\th_3}{\eta}\right)^4 + \left(\frac{\th_4}{\eta}\right)^4 -\left(\frac{\th_2}{\eta}\right)^4
\right|^2
\cdot \left| \frac{\th_3 \th_2}{\eta^2}\right|^4 \neq 0,
$$
because of the cancellation of the phase factors in the coefficient of the term $\dsp \left| \frac{\th_3 \th_2}{\eta^2}\right|^4$ 
appearing in \eqn{ex2 ab odd}.
In this way we conclude that  the criterion \eqn{criterion} is not satisfied in this case. 

~


\noindent
{\bf 3. 
Example from heterotic string on
$ T^4[D_4] \times S^1 \times S^1_{N\ep}$ :  }

The third example is a model from the heterotic string on 
$
\br^{3,1} \times T^4[D_4] \times S^1 \times S^1_{N\ep},
$
and 
\begin{equation}
g= (-1)^{F_R} \otimes s_R[4],
\label{g ex 3}
\end{equation}
where $s_R[4]$ acts on $T^4[D_4]$.
The total building blocks \eqn{hetero ab general} are written as 
\begin{align}
Z_{(a,b)}(\tau) & = 
\frac{1}{\tau_2 \left|\eta(\tau)\right|^4}\, 
Z^{S^1}(\tau) \cdot \overline{h_{(a,b)}} 
\cdot
  \ep^{[4]}_{(a,b)} \, 
\overline{\left(\tchi^{A_1}_{(a,b)}\right)^{4}} \, \chi^{D_4}_{(a,b)} \cdot \left[\chi^{E_8}_0\right]^2 .
\label{ex3 ab general}
\end{align}
Here, 
$\chi^{E_8}_0$ denotes the character of the basic representation of affine $E_8$ with level 1 given in \eqn{ch E8}.
We find that the sectors with $a\in 2\bz+1$, $b\in 2\bz$ satisfy the criterion \eqn{criterion} due to the phase factor $\ep^{[4]}_{(a,b)}$.
We thus obtain the vanishing cosmological constant $\dsp \lim_{\ep\, \rightarrow \, +0} \La_{\cM[\ep]} =0$,  as in the first example. 

Moreover, we obtain for the untwisted sector 
\begin{align}
\hspace{-0.5cm}
Z_0(\tau) 
& = \frac{1}{2}
\frac{1}{\tau_2 \left|\eta(\tau)\right|^4}\, 
Z^{S^1}(\tau) \cdot 
\overline{
\left[
\left(\frac{\th_3}{\eta}\right)^4 - \left(\frac{\th_4}{\eta}\right)^4 - \left(\frac{\th_2}{\eta}\right)^4
\right]
}
\nn
& \hspace{2cm} \times   
\frac{1}{2} \left[ \left|\frac{\th_3}{\eta}\right|^8 
+ \left|\frac{\th_4}{\eta}\right|^8
+ \left|\frac{\th_2}{\eta}\right|^8
\right]
\cdot\left[\chi^{E_8}_0 \right]^2
\nn
& + \frac{1}{2}
\frac{1}{\tau_2 \left|\eta(\tau)\right|^4}\, 
Z^{S^1}(\tau) \cdot 
\overline{
\left[
\left(\frac{\th_3}{\eta}\right)^4 - \left(\frac{\th_4}{\eta}\right)^4 +\left(\frac{\th_2}{\eta}\right)^4
\right]
}\,
\nn
& \hspace{2cm} \times  
\overline{\left(\frac{\th_3 \th_4}{\eta^2}\right)^2} \, 
\frac{1}{2} \left[\left(\frac{\th_3}{\eta}\right)^4 + \left(\frac{\th_4}{\eta}\right)^4\right]
\cdot\left[\chi^{E_8}_0 \right]^2 .
\label{ex3 a=0 b}
\end{align}
This does not vanish, meaning that we do not have the bose-fermi cancellation in the theory $\cM[\ep]$ again.  

As is familiar, there appear various non-abelian gauge symmetries in the heterotic string models, in contrast with the type II cases,  
which originate from the left-moving vertex operators with 
conformal weight $h=1$. 
In the present case the relevant orbifold action \eqn{g ex 3} does not affect such bosonic massless spectrum lying in 
$Z_0(\tau) $ and also the $S^1_{N\ep}$-sector omitted here.  
We then find the non-abelian gauge symmetry 
$U(1) \times U(1) \times SO(8)\times E_8 \times E_8$
(at generic points of the moduli space of the $S^1$-sector), while we always have an abelian gauge group $U(1)^6$ that originates from the right-mover. 
On the other hand,  no massless space-time fermions appear  
in the manner similar to  the first example. 

~


\noindent
{\bf 4. Example from heterotic string on
$ T^4[D_4] \times S^1 \times S^1_{N\ep}$ :  }

The fourth example is defined  for
the same background of example 3,  
$
\br^{3,1} \times T^4[D_4] \times S^1 \times S^1_{N\ep},
$
but we here take
\begin{equation}
g= (-1)^{F_R} \otimes s_L[4],
\label{g ex 4}
\end{equation}
instead of \eqn{g ex 3} as the orbifold action.
Then,
the total building blocks \eqn{hetero ab general} become 
\begin{align}
Z_{(a,b)}(\tau) & = 
\frac{1}{\tau_2 \left|\eta(\tau)\right|^4}\, 
Z^{S^1}(\tau) \cdot \overline{h_{(a,b)}} 
\cdot
  \ep^{[-4]}_{(a,b)} \, 
\left(\tchi^{A_1}_{(a,b)}\right)^{4} \, \overline{\chi^{D_4}_{(a,b)}} \cdot \left[\chi^{E_8}_0\right]^2 ,
\label{ex4 ab general}
\end{align}
which again satisfy the criterion \eqn{criterion}.

The untwisted partition function becomes
\begin{align}
\hspace{-0.5cm}
Z_0(\tau) 
& = \frac{1}{2}
\frac{1}{\tau_2 \left|\eta(\tau)\right|^4}\, 
Z^{S^1}(\tau) \cdot 
\overline{
\left[
\left(\frac{\th_3}{\eta}\right)^4 - \left(\frac{\th_4}{\eta}\right)^4 - \left(\frac{\th_2}{\eta}\right)^4
\right]
}
\nn
& \hspace{2cm} \times   
\frac{1}{2} \left[ \left|\frac{\th_3}{\eta}\right|^8 
+ \left|\frac{\th_4}{\eta}\right|^8
+ \left|\frac{\th_2}{\eta}\right|^8
\right]
\cdot\left[\chi^{E_8}_0 \right]^2
\nn
& + \frac{1}{2}
\frac{1}{\tau_2 \left|\eta(\tau)\right|^4}\, 
Z^{S^1}(\tau) \cdot 
\overline{
\left[
\left(\frac{\th_3}{\eta}\right)^4 - \left(\frac{\th_4}{\eta}\right)^4 +\left(\frac{\th_2}{\eta}\right)^4
\right]}\,
\nn
& \hspace{2cm} \times  
\left(\frac{\th_3 \th_4}{\eta^2}\right)^2 \, 
\frac{1}{2} \overline{ \left[\left(\frac{\th_3}{\eta}\right)^4 + \left(\frac{\th_4}{\eta}\right)^4\right] }
\cdot\left[\chi^{E_8}_0 \right]^2 ,
\label{ex4 a=0 b}
\end{align}
which again does not vanish. 

This time, the unbroken gauge symmetry originating from the left-mover is given as 
$U(1) \times U(1) \times SO(4)\times SO(4) \times E_8 \times E_8$, 
that is, the $SO(8)$ gauge symmetry is broken to $SO(4) \times SO(4)$ by the relevant orbifolding \eqn{g ex 4}. 
On the other hand, as opposed to the third example, we have massless space-time fermions which belong to the vector representation for
both factors of the $SO(4)$ gauge groups\footnote
  {The simplest way to observe these aspects is as follows;
The $T^4[D_4]$ sector is fermionized in the standard manner, that is, described by the 8 pairs of chiral fermions $(\la_L^i, \, \la_R^i)$ ($i=1,\ldots,8$),
and $s_L[4]$ is just regarded  as the sign-flip of the 4 left-moving fermions, say, $\la_L^i$, $i=1,\ldots, 4$.
Then, one can easily understand the massless spectrum mentioned here. 
}.


~

\section{Discussions and Comments}

In this paper we have demonstrated how one can systematically construct the {\em point particle theories}
that realize the vanishing one-loop cosmological constant without the bose-fermi cancellation, namely, 
the theories with the property \eqn{vanishing cc particle}.
The main idea to reach the desired theories is to utilize the building blocks of 
certain asymmetric orbifolds of supersymmetric string vacua, in which all the space-time supercharges are removed  at least in the untwisted sectors. 
We have interpreted 
the partition function for 
these untwisted sectors, 
which is not modular invariant, 
as  that  of the particle theories satisfying \eqn{vanishing cc particle}.
We have clarified the several conditions to be satisfied by the relevant orbifold models, 
and assumed the consistency of the  superstring theories on these orbifolds. 
Especially, we required the world-sheet superconformal symmetry in the RNS formalism.


However, 
if one focuses only on the spectrum,
it is possible to relax the consistency 
conditions of the original superstring theories, or even not to 
start from superstring theories.
In such cases, the possibility of the construction 
is largely enhanced, since appropriate modular forms
can be used as building blocks,
regardlessly of the consistency at intermediate stages. 


An important issue here is the consistency {\em after interactions are turned on.}
If the resultant particle theories are still consistent, 
we may adopt them, 
as our purpose is to construct the non-SUSY {\em particle theories\/}  
rather than the non-SUSY string vacua.
We have, however,  considered the particle theories
which descend from consistent superstring theories
in this paper,
since  the consistency of the latter 
may be inherited by the former.

At one loop, for example, the multi-particle scattering amplitudes must be compatible with the requirement of unitarity.
It would be possible to reduce this issue to that in superstring theory. 
Indeed, the superstring multi-particle amplitudes 
are given by the integral of 
the contributions from each spin structure over the torus modulus
 (see e.g. \cite{Green:1987mn,DHoker:1988pdl}). 
 Each contribution is
 invariant under the modular subgroup $\Gamma(2)$,  
which preserves the spin structures. Furthermore, 
it is decomposed into
the factors coming from the partition functions and from the 
vertex operators, and those factors are separately modular
covariant.
One may thus expect that the consistency
of the relevant amplitudes in the particle theories 
is deduced from
that of superstring 
by an argument similar to the one given in section \ref{sec 2},
with the help of the modular property of the amplitudes.
It would be also possible to confirm the UV-finiteness of the multi-particle amplitudes 
in our particle theories,
even though the integration region of the modulus (or Schwinger parameter) is 
$\cS$, rather than $\cF$, as for the cosmological constant.

Once establishing the consistency of multi-particle amplitudes at one-loop, one would be able to evaluate 
the higher loop corrections to cosmological constant by the `cutting and sewing procedure' of Riemann surfaces. 
We expect that the modular arguments presented in section \ref{sec 2} still work (especially, the natural extensions of  \eqn{eval La ep 1} to higher  genera).
If this is the case,  the higher loop cosmological constant  would also be shown to  behave as 
$$
\La_{\msc{higher loop}} \sim \ep^\lambda \times [\mbox{power of coupling constant}] \times  M_{\ep}^D
, 
$$
similarly to  the one-loop case, where $\lambda$ is a positive number depending on the loop number.
We would like to discuss these issues in more detail elsewhere.

~

Let us further add several comments:
\begin{description}
\item[(i)]
We have demonstrated the two types of constructions based on type II and heterotic strings by utilizing the building blocks 
studied in \cite{SSLie}. They might look quite similar. However, we have a crucial difference between them.

In the type II cases, the spectra read off from $Z_0(\tau)$ (and thus, the theory $\cM[\ep]$) do not include    
massless space-time fermions, 
while the heterotic cases can do. This aspect originates from the fact  that 
the chiral half-shift only acts on the bosonic coordinates. 
In fact, the possible massless fermions 
for the type II cases
should be of the form
$$
\ket{\mbox{R-vacuum}}_L \otimes \psi^i_{R,\,-1/2} \ket{0}_R, ~~ \mbox{or} ~~
 \psi^i_{L,\,-1/2} \ket{0}_L \otimes \ket{\mbox{R-vacuum}}_R .
$$
However, both of them are projected out by the orbifold action \eqn{g type II}, since 
the chiral half-shifts $s_L[*]$, $s_R[*]$ assign +1 to any NS massless states
in addition to $ (-1)^{F_L+F_R} = -1$.

On the other hand, in the heterotic models, one can obtain
 the massless fermionic states of the form
$$
\ket{p_L}_L \otimes \ket{\mbox{R-vacuum}}_R, 
$$
where $\ket{p_L}_L$ is any massless Fock vacuum ({\em i.e.} $h=1$) with $s_L[*]=-1$.
(See example 4 
presented in subsection \ref{subsec: ex}.)

~


\item[(ii)]
As mentioned above, the models we constructed satisfy the unitarity condition {\bf (3)}, 
that is, the unitarity 
in the untwisted sector 
with $a=0$ in $Z_{(a,b)}$, which is easily confirmed. 
We need not impose the same condition 
in the twisted sectors with $a\neq 0$
at least at the level of the spectrum, though the condition for $a\neq 0$ is also 
satisfied in our examples.

~


\item[(iii)]
We here only treated 
the twist for the world-sheet fermions 
by $(-1)^{F_L+F_R}$ in the type II cases and 
by $(-1)^{F_R}$ in  the heterotic cases
in order to break the space-time SUSY,
in other words, to achieve the condition {\bf (2)}, 
$Z_0(\tau) \neq 0$.
As another possibility to break SUSY, one can also make use of the chiral reflections 
$$
(X_L^i, \, X_R^i) ~ \longrightarrow (-X^i_L, \, X^i_R),
\hspace{1cm}
(\psi_L^i, \, \psi_R^i) ~ \longrightarrow (-\psi^i_L, \, \psi^i_R),
$$ 
or those for the right-mover.
The fermionic building blocks for the chiral reflection acting on the 
$2p \, (p=1,...,4)$ components $(\-_L)^{\otimes 2p}$ are explicitly given in 
appendix A and denoted as $g^{[p]}_{(a,b)}(\tau)$. They do not identically vanish for 
$p \neq 2$, 
which means the absence of  the bose-fermi cancellation. 

One can construct more elaborated 
models satisfying the requirements {\bf (1)} $\sim$ {\bf (5)} 
by combining the chiral half-shifts and the chiral reflections. 
The crucial point is that the total building blocks 
$Z_{(a,b)}(\tau)$ for the  odd sectors 
should include the non-trivial phase factors
such as $e^{\frac{i \pi}{8} K ab}$,
especially to satisfy the condition {\bf (5)}.
To this aim the inclusions of the chiral half-shifts are useful, since the phases that originate from the chiral reflections 
tend to be canceled out after combining the bosonic and fermionic building blocks. (Compare \eqn{ep r ab} and \eqn{gp ab}.)
We also note that the massless fermions can appear in these models  
{\em even in the type II cases}. This is in a sharp contrast with those constructed only with the chiral half-shifts 
and $(-1)^{F_L+F_R}$ mentioned in the comment {\bf (i)}.


\end{description}


~

\subsection*{Acknowledgments}
We would like to thank S. Iso for useful discussion.
This work is supported in part by 
JSPS KAKENHI Grant Number JP17K05406.


\newpage

\appendix


\section*{Appendix A: ~ Summary of Notations}

\setcounter{equation}{0}
\def\theequation{A.\arabic{equation}}

In appendix A, we summarize the notations used in this paper, 
and the building blocks to compose the modular invariants 
for the asymmetric orbifolds given in section \ref{sec:simple models}.
We set $q := e^{2\pi i \tau}$, $y := e^{2\pi i z}$ 
($\any \tau \in \bh^+$, $\any z \in \bc$).

~

\noindent
{\bf 1. Theta Functions} 
%
 \begin{align}
 & \dsp \th_1(\tau,z):=i\sum_{n=-\infty}^{\infty}(-1)^n q^{(n-1/2)^2/2} y^{n-1/2}
  \equiv  2 \sin(\pi z)q^{1/8}\prod_{m=1}^{\infty}
    (1-q^m)(1-yq^m)(1-y^{-1}q^m), \nn [-10pt]
   & \\[-5pt]
 & \dsp \th_2(\tau,z):=\sum_{n=-\infty}^{\infty} q^{(n-1/2)^2/2} y^{n-1/2}
  \equiv 2 \cos(\pi z)q^{1/8}\prod_{m=1}^{\infty}
    (1-q^m)(1+yq^m)(1+y^{-1}q^m), \\
 & \dsp \th_3(\tau,z):=\sum_{n=-\infty}^{\infty} q^{n^2/2} y^{n}
  \equiv \prod_{m=1}^{\infty}
    (1-q^m)(1+yq^{m-1/2})(1+y^{-1}q^{m-1/2}),  
\\
 &  \dsp \th_4(\tau,z):=\sum_{n=-\infty}^{\infty}(-1)^n q^{n^2/2} y^{n}
  \equiv \prod_{m=1}^{\infty}
    (1-q^m)(1-yq^{m-1/2})(1-y^{-1}q^{m-1/2}) . 
\\
& \Th{m}{k}(\tau,z):=\sum_{n=-\infty}^{\infty}\,
 q^{k(n+\frac{m}{2k})^2}y^{k(n+\frac{m}{2k})} ,
\\
& \tTh{m}{k}(\tau,z):=\sum_{n=-\infty}^{\infty}\, (-1)^n
 q^{k(n+\frac{m}{2k})^2}y^{k(n+\frac{m}{2k})} ,
\\
&
\eta(\tau) := q^{1/24}\prod_{n=1}^{\infty}(1-q^n).
 \end{align}
We often use abbreviations, $\th_i \equiv \th_i (\tau) \equiv \th_i(\tau, 0)$
 ($\th_1 \equiv \th_1(\tau)\equiv 0$), 
$\Th{m}{k} \equiv \Th{m}{k}(\tau) \equiv \Th{m}{k}(\tau,0)$, and 
$\tTh{m}{k} \equiv \tTh{m}{k}(\tau) \equiv \tTh{m}{k}(\tau,0)$.

%


~

\noindent
{\bf 2. Bosonic Building Blocks}

We next exhibit the bosonic building blocks 
for the `odd sectors' 
with $a\in 2\bz+1$ or $b \in 2\bz+1$
in the relevant asymmetric orbifolds repeatedly appearing  in section \ref{sec:simple models}.

~


%
%
%
%
%

\begin{itemize}
\item $A_1$-type :

We first describe the building blocks 
twisted by the `chiral reflection' $e^{i\pi J^1_0}$ in the current algebra $\widehat{A}_1$;
\begin{eqnarray}
\tchi^{A_1}_{(a,b)}(\tau) & := & 
\left\{
\begin{array}{ll}
\dsp 
\sqrt{
\frac{\th_3 \th_4}{\eta^2}
}
& ~~ (a\in 2\bz, ~ b\in 2\bz+1) ,\\
\dsp 
 \sqrt{
\frac{\th_3 \th_2}{\eta^2}
}
& ~~ (a\in 2\bz+1, ~ b\in 2\bz) ,\\
\dsp   \sqrt{
\frac{\th_4 \th_2}{\eta^2}
}
& ~~ (a\in 2\bz+1, ~ b\in 2\bz+1) .\\
\end{array}
\right.
\label{tchi A1 ab}
\end{eqnarray}
We note that the same functions can be rewritten in the expressions which look more natural for the `chiral half-shift' $e^{i \pi J^3_0}$; 
\begin{eqnarray}
\tchi^{A_1}_{(a,b)}(\tau) & \equiv & 
\left\{
\begin{array}{ll}
\dsp 
\frac{\tTh{0}{1}}{\eta} 
& ~~ (a\in 2\bz, ~ b\in 2\bz+1) ,\\
\dsp 
\sqrt{2} \frac{\Th{\frac{1}{2}}{1}}{\eta} 
& ~~ (a\in 2\bz+1, ~ b\in 2\bz) , \\
\dsp   
\sqrt{2} \frac{\tTh{\frac{1}{2}}{1}}{\eta}
& ~~ (a\in 2\bz+1, ~ b\in 2\bz+1) . \\
\end{array}
\right.
\label{tchi A1 ab 2}
\end{eqnarray}



We also introduce the functions
\begin{eqnarray}
\chi^{A_1}_{(a,b)}(\tau) & := & 
\left\{
\begin{array}{ll}
\dsp
\frac{1}{2} \left\{
\chi_+^{A_1}+ 
e^{- \frac{i\pi}{2}ab}
\chi_-^{A_1}
\right\}
& ~~ (a\in 2\bz, ~ b\in 2\bz+1) ,\\
\dsp
\frac{1}{\sqrt{2}} \left\{
\chi_0^{A_1}+ 
e^{\frac{i\pi}{2}ab}
\chi_1^{A_1}
\right\}
& ~~ (a\in 2\bz+1) .
\end{array}
\right.
\label{chi A1 ab}
\end{eqnarray}
where 
\begin{equation}
\chi^{A_1}_\ell(\tau) := \frac{\Th{\ell}{1}}{\eta}, 
\label{chi A1}
\end{equation}
denotes the spin $\ell/2$ character of $\widehat{A}_1$ with level 1 ($\ell=0, 1$), 
and we set 
\begin{equation}
\chi^{A_1}_{\pm}(\tau):= \chi_0^{A_1} \pm \chi_1^{A_1}.
\label{chi A1 pm}
\end{equation}

They yield the modular covariant  blocks 
\eqn{Z TXr ab} for the $T^1[A_1]$-case
with the phase factors $\ep^{[1]}_{(a,b)}$ defined in \eqn{ep r ab}, 
 that is, 
\begin{equation}
Z^{T^1[A_1]}_{(a,b)}(\tau) \equiv \ep^{[1]}_{(a,b)} \, \overline{\tchi^{A_1}_{(a,b)}(\tau)} \, \chi^{A_1}_{(a,b)}(\tau) 
\hspace{1cm}
(a \in 2\bz+1 , ~  \mbox{or ~}  b \in 2\bz+1) ,
\label{Z A1 ab}
\end{equation}
satisfies 
$$
\left. Z^{T^1[A_1]}_{(a,b)}(\tau) \right|_{T} =  Z^{T^1[A_1]}_{(a,a+b)}(\tau) ,
\hspace{1cm} 
\left. Z^{T^1[A_1]}_{(a,b)}(\tau) \right|_{S} =  Z^{T^1[A_1]}_{(b,-a)}(\tau) .
$$

~

\item $E_7$-type :

The relevant building blocks for the $E_7$-type are written as 
\begin{eqnarray}
\chi^{E_7}_{(a,b)}(\tau) & := & 
\left\{
\begin{array}{ll}
\dsp
\frac{1}{2} \left\{
\chi_+^{E_7}+ 
e^{\frac{i\pi}{2}ab}
\chi_-^{E_7}
\right\}
& ~~ (a\in 2\bz, ~ b\in 2\bz+1) , \\
\dsp
\frac{1}{\sqrt{2}} \left\{
\chi_0^{E_7}+ 
e^{- \frac{i\pi}{2}ab}
\chi_1^{E_7}
\right\}
& ~~ (a\in 2\bz+1),
\end{array}
\right.
\label{chi E7 ab}
\end{eqnarray}
where 
\begin{align}
& \chi^{E_7}_0(\tau) := \frac{1}{2\eta^7} \left[\Th{0}{1} \left(\th_3^6+\th_4^6\right)+ \Th{1}{1} \th_2^6\right],
\nn
& \chi^{E_7}_1(\tau) := \frac{1}{2\eta^7} \left[\Th{1}{1} \left(\th_3^6-\th_4^6\right)+ \Th{0}{1} \th_2^6\right],
\label{chi E7}
\end{align}
denotes the characters of  
the basic and the fundamental representation  of 
$\widehat{E}_7$ with level 1, which contain the states with dimension
$h=0$ and $h=\frac{3}{4}$  respectively, 
and we set 
\begin{equation}
\chi^{E_7}_{\pm}(\tau):= \chi_0^{E_7} \pm \chi_1^{E_7}.
\label{chi E7 pm}
\end{equation}

\eqn{chi E7 ab} yields the modular covariant blocks \eqn{Z TXr ab} for the $T^7[E_7]$-case
together with $\left(\tchi^{A_1}\right)^7$ and $\ep^{[7]}_{(a,b)}$.

~

\item
$D_r$-type :

The relevant building blocks for the $D_r$-type are written as 
\begin{eqnarray}
\chi^{D_r}_{(a,b)}(\tau) & := & 
\left\{
\begin{array}{ll}
\dsp
\frac{1}{2 \eta^r} \left\{
\th_3^r + 
e^{- \frac{i\pi r}{4}ab}
\th_4^r
\right\}
& ~~ (a\in 2\bz, ~ b\in 2\bz+1) ,\\
\dsp 
\frac{1}{2 \eta^r} \left\{
\th_3^r+ 
e^{\frac{i\pi r}{4}ab}
\th_2^r
\right\}
& ~~ (a\in 2\bz+1,
~ b\in 2\bz) ,\\
\dsp
\frac{1}{2 \eta^r} \left\{
\th_4^r+ 
e^{\frac{i\pi r}{4}ab}
\th_2^r
\right\}
& ~~ (a~ b \in 2\bz+1) . \\
\end{array}
\right.
\label{chi Dr ab}
\end{eqnarray}


We note that the modular covariant blocks for the $T^r[D_r]$-case are likewise composed  from 
the functions \eqn{chi Dr ab} as well as $\left(\tchi^{A_1}_{(a,b)}(\tau)\right)^r$, but with
{\em the phase factors slightly different from $\ep^{[r]}_{(a,b)}$};
\begin{equation}
Z^{D_r}_{(a,b)}(\tau) := \widetilde{\ep}^{[r]}_{(a,b)} \, \overline{\left(\tchi^{A_1}_{(a,b)}(\tau)\right)^r} \, \chi^{D_r}_{(a,b)}(\tau),
\label{Z Dr ab}
\end{equation}
where we define
\begin{equation}
\widetilde{\ep}^{[r]}_{(a,b)} := 
\left\{
\begin{array}{ll}
\left(\kappa_b\right)^r \ep^{[r]}_{(a,b)} 
& ~~ (a\in 2\bz, ~ b\in 2\bz+1) ,
\\
\left(\kappa_a \right)^r  \ep^{[r]}_{(a,b)} 
& ~~ (a \in 2\bz+1).
\end{array}
\right.
\end{equation}
The distinction between $\widetilde{\ep}^{[r]}_{(a,b)}$ and $\ep^{[r]}_{(a,b)}$  affects only in the cases of $r\in 2 \bz+1$,
and thus one may simply adopt $\ep^{[r]}_{(a,b)}$ in \eqn{Z TXr ab} as long as the number of `$D_{\msc{odd}}$-pieces' in the total Lie algebra lattice 
$X_r$ is even.
The building blocks for  odd $r $ extend those 
for even $r$  in \cite{SSLie}\footnote
  { In \cite{SSLie},  only the cases of $r \in 2\bz$ are treated among the  
$D_{r}$-type Lie algebra lattices, in which the phases $(\kappa_a)^r$ or $(\kappa_b)^r$  are absent. }.

\end{itemize}


The building blocks $Z^{T^r[X_r]}$ for the `even sectors'
 with $a, b \in 2\bz$
are given by the diagonal modular invariants 
of the characters of the associated affine Lie algebras.
Therefore, 
\begin{align}
 Z^{T^1[A_1]}(\tau)  
 &= \left| \chi_0^{A_1} \right|^2 + \left| \chi_1^{A_1} \right|^2
 \, , \qquad
  Z^{T^7[E_7]}(\tau)  
 = \left| \chi_0^{E_7} \right|^2 + \left| \chi_1^{E_7} \right|^2 
 \, , \nonumber \\
 Z^{T^r[D_r]}(\tau)  
 & = \left| \chi_0^{D_r} \right|^2 + \left| \chi_v^{D_r} \right|^2
 + \left| \chi_s^{D_r} \right|^2  + \left| \chi_c^{D_r} \right|^2 \, ,
\end{align}
where $\chi_{\ell}^{A_1}$ and $\chi_{\ell}^{E_7}$ are given in \eqn{chi A1}, \eqn{chi E7}, and for the $D_r$-cases, 
\begin{align}
& \chi_{0}^{D_r}  = \frac{1}{2\eta^r} \left( \theta_3^r + \theta_4^r\right)
\hspace{1cm} \mbox{(basic rep.)} ,
\nn
& \chi_{v}^{D_r}  = \frac{1}{2\eta^r} \left( \theta_3^r - \theta_4^r\right)
\hspace{1cm} \mbox{(vector rep.)} ,
\nn
& \chi_{s}^{D_r}  = \chi_{c}^{D_r} = \frac{1}{2\eta^r} \theta_2^r 
\hspace{1cm} \mbox{(spinor or cospinor rep.)}.
\end{align}


~

\noindent
We also note the familiar formula of the character of the basic representation of affine $E_8$ with level 1, 
\begin{equation}
\chi^{E_8}_0(\tau) = \frac{1}{2} \left[
\left(\frac{\th_3}{\eta}\right)^8 + \left(\frac{\th_4}{\eta}\right)^8 + \left(\frac{\th_2}{\eta}\right)^8
\right].
\label{ch E8}
\end{equation}

~


\noindent
{\bf 3. Fermionic Building Blocks }

Here we summarize the modular covariant chiral blocks of 
the world-sheet fermions appearing in section 3. 
%
%
We first describe those twisted by $(-1)^{F_L}$, where $F_L$ denotes the space-time fermion number operator,
\begin{eqnarray}
&& 
 h_{(a,b)}(\tau) \equiv \left\{
\begin{array}{ll}
\left(\frac{\th_3}{\eta}\right)^4
- \left(\frac{\th_4}{\eta}\right)^4
+ \left(\frac{\th_2}{\eta}\right)^4
& ~~ (a\in 2\bz, ~ b \in 2\bz+1) , \\
\left(\frac{\th_3}{\eta}\right)^4
+ \left(\frac{\th_4}{\eta}\right)^4
- \left(\frac{\th_2}{\eta}\right)^4
& ~~ (a\in 2\bz+1, ~ b \in 2\bz) , \\
-\left\{\left(\frac{\th_3}{\eta}\right)^4
+ \left(\frac{\th_4}{\eta}\right)^4
+ \left(\frac{\th_2}{\eta}\right)^4 \right\}
& ~~ (a\in 2\bz+1, ~ b \in 2\bz+1) , \\
\left(\frac{\th_3}{\eta}\right)^4
- \left(\frac{\th_4}{\eta}\right)^4
- \left(\frac{\th_2}{\eta}\right)^4
& ~~ (a\in 2\bz, ~ b \in 2\bz)  .
\end{array}
\right. 
\end{eqnarray}
The block for the `even sectors'  with $a,b \in 2\bz$ is 
the supersymmetric 
one that is identically zero. 
The modular covariance of 
$h_{(a,b)}(\tau)$
means that
\begin{align}
& h_{(a,b)}(\tau)|_T 
= - e^{-2\pi i \frac{1}{6}} h_{(a,a+b)}(\tau) ,
\hspace{1cm}
 h_{(a,b)}(\tau)|_S 
= h_{(b,-a)}(\tau).
\label{mc h ab}
\end{align}

~


We next consider the chiral reflection acting on the $2p \, (p=1, ...,4)$ components of the world-sheet fermions $\left(\-_L\right)^{\otimes 2p}$.
The relevant blocks are written as  
\begin{eqnarray}
g^{[p]}_{(a,b)}(\tau) 
&\equiv &
\left\{
\begin{array}{ll}
e^{- \frac{i \pi p}{4} ab}
\left\{
\left(\frac{\th_3}{\eta}\right)^{4-p}
\left(\frac{\th_4}{\eta}\right)^p
- 
e^{\frac{i \pi p}{2} ab}
\left(\frac{\th_4}{\eta}\right)^{4-p}
\left(\frac{\th_3}{\eta}\right)^p
\right\}
& ~~ (a\in 2\bz, ~ b\in 2\bz+1) ,\\
e^{\frac{i \pi p}{4} ab}
\left\{
\left(\frac{\th_3}{\eta}\right)^{4-p}
\left(\frac{\th_2}{\eta}\right)^p
- 
e^{- \frac{i \pi p}{2} ab}
\left(\frac{\th_2}{\eta}\right)^{4-p}
\left(\frac{\th_3}{\eta}\right)^p
\right\}
& ~~ (a\in 2\bz+1, ~ b\in 2\bz) , \\
- 
e^{ \frac{i \pi p}{4} ab}
\left\{
\left(\frac{\th_4}{\eta}\right)^{4-p}
\left(\frac{\th_2}{\eta}\right)^p
+ 
e^{- \frac{i \pi p}{2} ab}
\left(\frac{\th_2}{\eta}\right)^{4-p}
\left(\frac{\th_4}{\eta}\right)^p
\right\}
& ~~ (a\in 2\bz+1, ~ b\in 2\bz+1)  , \\
 \left(\frac{\th_3}{\eta}\right)^4
- \left(\frac{\th_4}{\eta}\right)^4
-\left(\frac{\th_2}{\eta}\right)^4
 & ~~ (a \in 2\bz, ~ b\in 2\bz) . 
\end{array}
\right.
\nn
&&
\label{gp ab}
\end{eqnarray}
The blocks with $p=2$ 
are supersymmetric and identically vanish. 
They satisfy the relation of the modular covariance as in \eqn{mc h ab}, that is,
\begin{align}
& g^{[p]}_{(a,b)}(\tau)|_T 
= - e^{-2\pi i \frac{1}{6}} g^{[p]}_{(a,a+b)}(\tau) ,
\hspace{1cm}
 g^{[p]}_{(a,b)}(\tau)|_S 
= g^{[p]}_{(b,-a)}(\tau).
\label{mc gp ab}
\end{align}
We note that the non-trivial phase factors are necessary to achieve the correct modular covariance in contrast to the previous ones 
$h_{(a,b)}(\tau)$.


~

\section*{Appendix B: ~ Free Compact Boson with Self-Dual Radius}

\setcounter{equation}{0}
\def\theequation{B.\arabic{equation}}

In appendix B, we summarize the basic fact about the $\widehat{A_1}$-symmetry with level 1 realized 
by the free  boson compactified on the circle with the self-dual radius $R = \sqrt{\al'}$.
Assuming the standard normalization of free boson $\dsp X_L(z)X_L(0) \sim - \frac{\al'}{2} \ln z$, 
the chiral part of the $A_1$-current algebra with level 1 is 
given by the currents,
\begin{align}
J^3_L(z) = \frac{1}{\sqrt{\al'}} i \del X_L (z), 
\hspace{1cm} J^{\pm}_L(z) \equiv J^1_L(z) \pm i J^2_L(z) = e^{\pm i \frac{2}{\sqrt{\al'}} X_L(z)}.
\end{align}
%
%
We note
that $e^{i \pi J^1_{L,0} } $ acts as the chiral reflection, 
\begin{align}
e^{i \pi J^1_{L,0} } \, : \,  X_L \, \longrightarrow \, - X_L,
\label{J1 reflection}
\end{align}
while $e^{i \pi J^3_{L,0} } $ acts as the `chiral half-shift'\footnote
   {Combining the left and right movers, 
$e^{i\pi J^3_{L,0}} \otimes e^{i\pi J^3_{R,0}}$ acts on the free boson 
$X\equiv X_L+X_R$
as $X \, \rightarrow \, X + \pi \sqrt{\al'}$, which is identified as 
the half-shift. 
}, 
\begin{align}
e^{i \pi J^3_{L,0} } \, : \,  X_L \, \longrightarrow \, X_L + \frac{\pi }{2}\sqrt{\al'}.
\label{J3 half-shift}
\end{align}



\end{document}